\documentclass[12pt,twosided]{article}
\date{}
\usepackage{extsizes}
\usepackage{amsmath}
\usepackage{amsfonts}
\usepackage{amssymb}
\usepackage{graphicx}
\usepackage{braket}
\date{\today}
\begin{document}

\title{{\bf{Exact solutions of a damped harmonic oscillator in a time dependent noncommutative space }}}

\author{
{\bf {\normalsize Manjari Dutta}$^{a}
$\thanks{manjaridutta@boson.bose.res.in}},
{\bf {\normalsize Shreemoyee Ganguly}
$^{b}$\thanks{ganguly.shreemoyee@gmail.com}},
{\bf {\normalsize Sunandan Gangopadhyay}
$^{c}$\thanks{ sunandan.gangopadhyay@bose.res.in, sunandan.gangopadhyay@gmail.com}}
\\
$^{a,c}$ {\normalsize Department of Theoretical Sciences},\\
{\normalsize S.N. Bose National Centre for Basic Sciences},\\
{\normalsize JD Block, Sector III, Salt Lake, Kolkata 700106, India}\\
$^{b}$ {\normalsize Department of Basic Science and Humanities,}\\
{\normalsize University of Engineering and Management (UEM),}\\
{\normalsize B/5, Plot No.III, Action Area-III, Newtown, Kolkata 700156}
%$^{c}${\normalsize National Institute for Theoretical Physics (NITheP), University of Stellenbosch, Stellenbosch 7600, South Africa}\\
%$^{d}${\normalsize Institute of Theoretical Physics, University of Stellenbosch, Stellenbosch 7600, South Africa}
}
\date{}

\maketitle
\begin{abstract}
\noindent In this paper we have obtained the exact eigenstates of a two dimensional damped harmonic oscillator in time dependent noncommutative space. It has been observed that for some specific choices of the damping factor and the time dependent frequency of the oscillator, there exists interesting solutions of the time dependent noncommutative parameters following from the solutions of the Ermakov-Pinney equation. Further, these solutions enable us to get exact analytic forms for the phase which relates the eigenstates of the Hamiltonian with the eigenstates of the Lewis invariant. We then obtain expressions for the matrix elements of the coordinate operators raised to a finite arbitrary power. From these general results we then compute the expectation value of the Hamiltonian. The expectation values of the energy are found to vary with time for different solutions of the Ermakov-Pinney equation corresponding to different choices of the damping factor and the time dependent frequency of the oscillator.

\end{abstract}

\vskip 1cm
%%%%%%%%%%%%%%%%%%%%%%%%%%%%%%%%%%%%%%%%%%%%%%%%%%%%%%%%%%%%%%%%%%
\newpage
\section{Introduction}
The study of time dependent classical as well as quantum harmonic oscillators 
has appealed to theoretical physicists since time immemorial. In the literature 
the work by Lewis {\it et al.}~\cite{Lewis} has lead to an upsurge of analysis 
of the Hamiltonian for the time dependent quantum harmonic oscillator using a 
class of exact invariants designed for such systems~\cite{Lewis2,Lewis3}. The problem 
becomes even more fascinating when one has a system of two such oscillators in 
two-dimensional space. Now, in order to address practical situations one needs 
to include damping in the system. Although there are several studies on the 
one-dimensional damped quantum harmonic oscillator in the past~~\cite{Sebawe}-\cite{Pedrosa3}, it's two-dimensional equivalent 
is a less explored system~\cite{Gouba}. The work by Lawson {\it et.al.}~\cite{Gouba} is 
one of the very few which analyses a two-dimensional damped quantum harmonic oscillator 
system. The solutions obtained by them for the mentioned system provides a 
platform to explore the construction of various coherent states with 
intriguing properties. 

In the present work we extend the study by Lawson 
{\it et.al.}~\cite{Gouba} and consider the two-dimensional damped quantum harmonic 
oscillator in noncommutative (NC) space. It has been argued that study of quantum mechanical systems in NC space is essential to ensure the attainment of 
gravitational stability~\cite{Doplicher} in the present theories of quantum 
gravity, namely, string theory \cite{amati, sw} and loop quantum gravity \cite{rov}. 
The simplest quantum mechanical setting in two dimensional NC space
consists of replacing the standard set of commutation relations between the
canonical coordinates by NC commutation relations $[X, Y]=i\theta$, where $\theta$ is a positive real constant. Quantum mechanical systems in such spaces have been studied extensively in the literature \cite{suss}-\cite{fgs}.
The study of a two-dimensional quantum harmonic oscillator in NC space with time dependent NC parameters was done in \cite{Dey}. However, their system was an undamped oscillator. The parametrized form of solutions obtained there offered an interesting possibility for study of generalized version of Heisenberg's 
uncertainty relations. Quantum damped harmonic oscillator 
on noncommuting two-dimensional space was studied in \cite{anto} where the exact propagator of the system was obtained and the thermodynamic properties of the system was investigated using the standard canonical density matrix.

In this work, a two-dimensional damped quantum harmonic oscillator in NC space
is considered once again. However, our focus of study is different than the work carried out in \cite{anto}. We first construct the Hamiltonian and then express it in terms of standard commutative variables. This is done in Section 2. Then we solve the Hamiltonian using the method of invariants~\cite{Lewis} and obtain the corresponding eigenfunction in 
Section 3. In doing so, although we start with the Hamiltonian and corresponding 
invariant in Cartesian coordinates, eventually we transform our operators to 
polar coordinates (following closely the procedure suggested in \cite{Dey}) for ease of solution. The form of the Lewis invariant in Cartesian coordinates with a Zeeman term in the Hamiltonian is an interesting result in itself and it also makes it easier to make a transition to it's polar form. It is to be noted that the eigenfunction 
of the Hamiltonian is a product of the eigenfunction of the invariant and a 
phase factor. Both the eigenfunction and phase factor are expressed in terms 
of time dependent parameters which obey the non-linear differential equation known as Ermakov-Pinney (EP) equation~\cite{Ermakov,Pinney}. 
%The phase is obtained in an integral form.
Next, in Section 4 we judiciously choose the parameters of the damped system 
such that they satisfy all the equations representing the system as well as 
provide us with an exact closed form solution of the Hamiltonian. The 
solutions of the NC parameters obtained in our analysis turns out to be such that the phase factor in an integral form given in \cite{Dey} is exactly integrable for various kinds of dissipation. Then 
in Section 5 we device a procedure to calculate the matrix element of a finite arbitrary power of the position operator with respect to the exact solutions 
for Hamiltonian eigenstates. Using these expressions we proceed to calculate 
the expectation value of energy and study the 
evolution of the energy expectation value of the system with time 
for various types of damping. In Section 6 we summarize our results.

%%%%%%%%%%%%%%%%%%%%%%%%%%%%%%%%%%%%%%%%%%%%%%%%%%%%%%%%%%%%%%%%%%%%%%%%%%%%%%%

%%%%%%%%%%%%%%%%%%%%%%%%%%%%%%%%%%%%%%%%%%%%%%%%%%%%%%%%%%%%%%%%%%%%%%%%%%%%%%%

\section{Model of the two-dimensional harmonic oscillator}
The system we consider is a combination of two non-interacting damped harmonic oscillators in two dimensional NC space. The oscillators have equal time dependent 
frequencies, time dependent coefficients of friction and equal mass in 
NC space. Such a model of damped harmonic oscillator was 
considered in an earlier communication~\cite{Gouba} in commutative space. In this work, we extend the model by considering the system in NC space\footnote{We shall be considering NC phase space in our work. However, we shall generically refer this as NC space.}.

\noindent The Hamiltonian of the system has the following form,
\begin{equation}
H(t)=\dfrac{f(t)}{2M}({P_1}^2+{P_2}^2)+\dfrac{M\omega^2(t)}{2f(t)}({X_1}^2+{X_2}^2)\label{1}
\end{equation}
where the damping factor $f(t)$ is given by,
\begin{equation}
f(t)=e^{-\int_{0}^t\eta(s)ds} 
\label{1x}
\end{equation}
with $\eta(s)$ being the coefficient of friction. Here $\omega(t)$ is the 
time dependent angular frequency of the oscillators and $M$ is their mass.
It should be noted that in commutative space, the model with $f(t)=e^{-\Gamma t}$ and $\omega(t)=\omega_0$, with $\Gamma$ and $\omega_0$
being positive constants, is said to be the two-dimensional Caldirola and Kanai Hamiltonian \cite{caldi, kanai}.
The position and momentum 
coordinates $(X_i,P_i)$ are noncommuting variables in NC space, that is, 
their commutators are $[X_1,X_2]~\neq~0$ and $[P_1,P_2]~\neq~0$. The 
corresponding canonical variables $(x_i,p_i)$ in commutative space are such 
that the commutator $[x_i,p_j]=i\hbar\delta_{i,j}$, $[x_i,x_j]=0=[p_i,p_j]$; ($i,j=1,2$).

In order to express the NC Hamiltonian in terms of the standard commutative variables explicitly, we apply the standard Bopp-shift relations \cite{mez} ($\hbar=1$): 
\begin{eqnarray}
& X_1=x_1-\dfrac{\theta(t)}{2}p_2\,\,\,;\,\,\,X_2=x_2+\dfrac{\theta(t)}{2}p_1\\
& P_1=p_1+\dfrac{\Omega(t)}{2}x_2\,\,\,;\,\,\,P_2=p_2-\dfrac{\Omega(t)}{2}x_1 \,\,.
\label{eqn1}
\end{eqnarray}
Here $\theta(t)$ and $\Omega(t)$ are the NC parameters for space 
and momentum respectively, such that $[X_1,X_2]~=i\theta(t)$, 
$[P_1,P_2]~=i\Omega(t)$ and $[X_1,P_1]=i[1+\frac{\theta(t)\Omega(t)}{4}]=[X_2,P_2]$; ($X_1\equiv X$, $X_2 \equiv Y$, $P_1\equiv P_x$, $P_2 \equiv P_y$). 

\noindent The Hamiltonian in terms of $(x_i,p_i)$ coordinates is therefore given by the following 
relation,
\begin{equation}
H=\dfrac{a(t)}{2}({p_1}^2+{p_2}^2)+\dfrac{b(t)}{2}({x_1}^2+{x_2}^2)+c(t)({p_1}{x_2}-{p_2}{x_1})\,\,\,.\label{eqn2}
\end{equation}
The time dependent coefficients in the above Hamiltonian are given as,
\begin{eqnarray}
a(t)&=&\dfrac{f(t)}{M}+\dfrac{M{\omega^2(t)}\theta^2(t)}{4f(t)}\label{3} \\
 b(t)&=&\dfrac{f(t){\Omega^2(t)}}{4M}+\dfrac{M{\omega^2(t)}}{f(t)}\label{4} \\
 c(t)&=&\dfrac{1}{2}\left[\dfrac{f(t)\Omega(t)}{M}+\dfrac{M\omega^2(t)\theta(t)}{f(t)} \right]. 
%&=&\left[\omega(t)\sqrt{\dfrac{M\,a(t)}{f(t)}-1}+\sqrt{\dfrac{b(t)\,f(t)}{M}-\omega^2(t)} \right].
\label{eqn3}
\end{eqnarray}
%\textbf{Substituting the form of the NC parameters from the Eqn(s).(\ref{3},\ref{4}), an other form of $c(t)$ in terms of frequency and the damping factor is found as}
%\begin{equation}
%c(t)=\left[\omega(t)\sqrt{\dfrac{M\,a(t)}{f(t)}-1}+\sqrt{\dfrac{b(t)\,f(t)}{M}-\omega^2(t)} \right]\label{ct}
%\end{equation}
Here it must be noted that although our Hamiltonian given by Eqn.(\ref{eqn2}) has the same form as that in 
\cite{Dey} to study a system of a two dimensional harmonic oscillator in NC space, the time dependent 
Hamiltonian coefficients (given by Eqn(s).(\ref{eqn3})) have very different form. This is because our system is that of 
a damped harmonic oscillator in two-dimensional NC space. Thus, the damping factor $f(t)$ modulates and alters 
the Hamiltonian coefficients from the form considered in earlier study \cite{Dey}.

%%%%%%%%%%%%%%%%%%%%%%%%%%%%%%%%%%%%%%%%%%%%%%%%%%%%%%%%%%%%%%%%%%%%%%%%%%%%%%%%%%%%%%
%%%%%%%%%%%%%%%%%%%%%%% SOLUTION %%%%%%%%%%%%%%%%%%%%%%%%%%%%%%%%%%%%%%%%%%%%%%%%%%%%%

\section{Solution of the model Hamiltonian}
In order to find the solutions of the model Hamiltonian $H(t)$ (Eqn.(\ref{eqn2}))
representing the two-dimensional damped harmonic oscillator in 
NC{ space, we follow the route suggested by Lewis {\it et.al.}~\cite{Lewis} in their work. First we 
construct the time-dependent Hermitian invariant operator $I(t)$ corresponding to our Hamiltonian operator $H(t)$ 
(given by Eqn.(\ref{eqn2})). This is because if one can solve for the eigenfunctions of $I(t)$, $\phi(x_1,x_2)$, such 
that,
\begin{equation}
I(t)\phi(x_1,x_2)=\epsilon \phi(x_1,x_2)
\label{eqnegn}
\end{equation}
where $\epsilon$ is an eigenvalue of $I(t)$ corresponding to eigenstate $\phi(x_1,x_2)$, one can obtain the 
eigenstates of $H(t)$, $\psi(x_1,x_2,t)$, using the relation given by Lewis {\it et. al.}~\cite{Lewis} which is as 
follows, 
\begin{equation}
\psi(x_1,x_2,t)=e^{i\Theta(t)}\phi(x_1,x_2)
\label{eqnpsi}
\end{equation}
where the real function $\Theta(t)$ which acts as the phase factor will be discussed in details later. 

%%%%%%%%%%%%%%%%%%%%%%%%%%%%%%%%%%%%%%%%%%%%%%%%%%%%%%%%%%%%%%%%%%%%%%%%%%%%%%%

%%%%%%%%%%%%%%%%%%%%%%%%%%%%%%%%%%%%%%%%%%%%%%%%%%%%%%%%%%%%%%%%%%%%%%%%%%%%%%%

\subsection{The Time Dependent Invariant}
Next, following the approach taken by Lewis {\it et.al.}~\cite{Lewis}, we need to construct the operator $I(t)$ which 
is an invariant with respect to time, corresponding to the Hamiltonian $H(t)$, as mentioned earlier, such 
that $I(t)$ satisfies the condition,
\begin{equation}
\dfrac{dI}{dt}=\partial_t{I}+\dfrac{1}{i}[I,H]=0.
\label{eqn4}
\end{equation}
The procedure is to choose the Hermitian invariant $I(t)$ to be of the same homogeneous quadratic form defined by Lewis 
{\it et. al.}~\cite{Lewis} for time-dependent harmonic oscillators. However, since we are dealing with a 
two-dimensional system in the present study, $I(t)$ takes on the following form,
\begin{equation}
I(t)=\alpha(t)({p_1}^2+{p_2}^2)+\beta(t)({x_1}^2+{x_2}^2)+\gamma(t)(x_1{p_1}+p_2{x_2}).
\label{eqn5}
\end{equation}
Here we will consider $\hbar=1$ since we choose to work in natural units. Now, using the form of $I(t)$ defined by 
Eqn.(\ref{eqn5}) in  Eqn.(\ref{eqn4}) and equating the coefficients of the canonical variables, we get the 
following relations,
\begin{eqnarray}
\dot{\alpha}(t)&=&-a(t)\gamma(t)\label{eqn6}\\
\dot{\beta}(t)&=&b(t)\gamma(t)\label{eqn7}\\
\dot{\gamma}(t)&=&2\left[\,b(t)\alpha(t)-\beta(t)a(t)\,\right]
\label{eqn8}
\end{eqnarray}
where dot denotes derivative with respect to time $t$.

\noindent To express the above three time dependent parameters $\alpha$,$\beta$ and $\gamma$ in terms of a single time 
dependent parameter, we parametrize $\alpha(t)=\rho^{2}(t)$. Substituting this in Eqn(s).(\ref{eqn6}, \ref{eqn8}), we 
get the other two parameters in terms of $\rho(t)$ as, 
\begin{eqnarray}
\gamma(t)&=&-\dfrac{2\rho\dot{\rho}}{a(t)}\label{eqn9}\\
\beta(t)&=&\dfrac{1}{a(t)}\left[\dfrac{{\dot{\rho}^2}}{a(t)}+{{\rho}^2}b+\dfrac{\rho\ddot{\rho}}{a(t)}-\dfrac{\rho\dot{\rho}\dot{a}}{a^2} \right].\label{eqn10}
%\beta(t)&=&\dfrac{1}{a(t)}\left[\dfrac{{\dot{\rho}^2}}{a(t)}+\dfrac{{\xi^2}{a(t)}}{\rho^2} \right]\label{eqn10}
\end{eqnarray}
Now, substituting the value of $\beta$ in Eqn.(\ref{eqn7}), we get a non-linear equation in 
$\rho(t)$ which has the form of the non-linear Ermakov-Pinney (EP) equation with a dissipative 
term~\cite{Dey, Ermakov, Pinney}. The form of the non-linear equation is as follows, 
\begin{equation}
\ddot{\rho}-\dfrac{\dot{a}}{a}\dot{\rho}+ab\rho={\xi^2}\dfrac{a^2}{\rho^3}~.
\label{eqn11}
\end{equation} 
where ${\xi^2}$ is a constant of integration.  This equation has similar form to the EP equation obtained in \cite{Dey}, which is expected since 
our $H(t)$ has the same form as theirs. However, once again we should recall the fact that the explicit form of the time-dependent 
coefficients are different due to the presence of damping.

\noindent Now, using the EP equation we get a simpler form of $\beta$ as,
\begin{eqnarray}
\beta(t)&=&\dfrac{1}{a(t)}\left[\dfrac{{\dot{\rho}^2}}{a(t)}+\dfrac{{\xi^2}{a(t)}}{\rho^2} \right].
\label{eqnew}
\end{eqnarray}  
\noindent Next, substituting the expressions of $\alpha$, $\beta$ and $\gamma$ in 
Eqn.(\ref{eqn5}), we get the following 
expression for $I(t)$,
\begin{equation}
I(t)=\rho^2({p_1}^2+{p_2}^2)+\left(\dfrac{\dot{\rho}^2}{a^2}+\dfrac{{\xi^2}}{\rho^2}\right)({x_1}^2+{x_2}^2)-\dfrac{2\rho\dot{\rho}}{a}(x_1{p_1}+p_2{x_2}).
\label{eqn12}
\end{equation}
The form of the Lewis invariant in Cartesian coordinates will be used later to go over to it's polar coordinate form.
The solution of the EP equation under various physically significant conditions shall be discussed later.

%%%%%%%%%%%%%%%%%%%%%%%%%%%%%%%%%%%%%%%%%%%%%%%%%%%%%%%%%%%%%%%%%%%%%%%%%%%%%%%%%%%%%%%%%%%%%%%%%%%%%%%%%%%%%%%%%%

%%%%%%%%%%%%%%%%%%%%%%%%%%%%%%%% Ladder Operators %%%%%%%%%%%%%%%%%%%%%%%%%%%%%%%%%%%%%%%%%%%%%%%%%%%%%%%%%%%%%%%%

\subsection{Construction of Ladder operators} 
Now that we have the required Hermitian invariant $I(t)$, we proceed to calculate it's eigenstates using the operator approach. For 
this purpose we need to first construct some ladder operators. To do this, we first need to transform the form of $I(t)$ 
(given by Eqn.(\ref{eqn12})) to a more manageable form. For this we invoke a unitary transformation using a suitable unitary 
operator $\hat{U}$ having the following form, 
\begin{eqnarray}
\hat{U}=exp\left[-\dfrac{i\dot{\rho}}{2a(t)\rho}({x_1}^2+{x_2}^2)\right],\,\,\,
\hat{U^{\dagger}}\hat{U}=\hat{U}\hat{U^{\dagger}}=\textbf{I}.
\label{eqn13}
\end{eqnarray}
Defining, 
\begin{eqnarray}
\phi^{'}(x_1,x_2)=\hat{U}\phi(x_1,x_2) \,\,\,,\,\,\,
I^{'}(t)&=\hat{U}I\hat{U^\dagger}\,\,\;
\label{eqn14}
\end{eqnarray}
where $\phi(x_1,x_2)$ is an eigenfunction of $I(t)$ as introduced in Eqn.(\ref{eqnegn}), 
then, using Eqn(s).(\ref{eqnegn},\ref{eqn14}), we get, 
\begin{align}
I^{'}\phi^{'}=\hat{U}I\hat{U^\dagger}\hat{U}\phi=\hat{U}I\phi=\hat{U}\epsilon\phi=\epsilon\phi^{'}.
\end{align}
The transformed expression of the invariant, $I^{'}(t)$, using Eqn.(\ref{eqn14}), has the following form, 
\begin{align}
I^{'}(t)=\rho^2({p_1}^2+{p_2}^2)+\dfrac{{\xi^2}}{\rho^2}({x_1}^2+{x_2}^2)\,\,.
\label{eqn15}   
\end{align}
This transformed form of the invariant, $I^{'}(t)$, has exactly the same form as that of the Hamiltonian for a time dependent 
two-dimensional simple harmonic oscillator. So, we can introduce the corresponding ladder operators for $\hat{I^{'}}(t)$ to be 
given by,
\begin{eqnarray}
{\hat{a}_j}^{'}=\dfrac{1}{\sqrt{2\xi}}\left(\dfrac{\xi}{\rho}{\hat{x}}_j+i\rho{\hat{p}}_j\right)\,\,\,,\,\,\,{{\hat{a}_j}^{'\dagger}}=\dfrac{1}{\sqrt{ 2\xi}}\left(\dfrac{\xi}{\rho}{\hat{x}}_j-i\rho{\hat{p}}_j\right)
\label{eqn16}
\end{eqnarray}
where $j=1,2$ and the operators satisfy the commutation relation $[{{\hat{a}_i}^{'}},{{\hat{a}_j}^{'\dagger}}]=\delta_{ij}$.

\noindent Now we make the reverse transformation to get the expression of the unprimed ladder operators:
\begin{eqnarray}
\hat{a_j}(t)&=&{\hat{U}}^\dagger{\hat{a}_j}^{'}\hat{U}=\dfrac{1}{\sqrt{2\xi}}\left[\dfrac{\xi}{\rho}x_j+i\rho{p_j}-\dfrac{i\dot{\rho}}{a(t)}x_j\right]\\
\label{eqn17}
\hat{{a_j}^{\dagger}}(t)&=&{\hat{U}}^\dagger{{\hat{a}_j}^{'\dagger}}\hat{U}=\dfrac{1}{\sqrt{2\xi}}\left[\dfrac{\xi}{\rho}x_j-i\rho{p_j}+\dfrac{i\dot{\rho}}{a(t)}x_j\right].
\label{eqn18} 
\end{eqnarray}
It can be easily checked using the algebra of the primed ladder operators that $[{{\hat{a}_i}},{{\hat{a}_j}^{\dagger}}]=\delta_{ij}$.

\noindent We now set $\xi=1$ and consider two linear combinations of the above two operators such that,
\begin{eqnarray}
\hat{a}(t)=-\dfrac{i}{\sqrt{2}}(\hat{a}_1-i\hat{a}_2)
=\dfrac{1}{2}\left[\rho(\hat{p_1}-i\hat{p_2})-\left(\dfrac{i}{\rho}+\dfrac{\dot{\rho}}{a(t)} \right)(\hat{x_1}-i\hat{x_2})\right]
\label{eqn19}
\end{eqnarray}
and
\begin{eqnarray}
{\hat{a}}^\dagger(t)=\dfrac{i}{\sqrt{2}}({\hat{a}_1}^\dagger+i{\hat{a}_2}^\dagger)
=\dfrac{1}{2}\left[\rho(\hat{p_1}+i\hat{p_2})+\left(\dfrac{i}{\rho}-\dfrac{\dot{\rho}}{a(t)} \right)(\hat{x_1}+i\hat{x_2})\right].
\label{eqn20}
\end{eqnarray}
These also satisfy the commutation relation $[\hat{a},\hat{a}^\dagger]=1$.

%%%%%%%%%%%%%%%%%%%%%%%%%%%%%%%%%%%%%%%%%%%%%%%%%%%%%%%%%%%%%%%%%%%%%%%%%%%%%%%%%%%%%%%%%%%%%%%%%%%%%

%%%%%%%%%%%%%%%%%%%%%%%%%%%%%%%%%%%%%%%%%%%%%%%%%%%%%%%%%%%%%%%%%%%%%%%%%%%%%%%%%%%%%%%%%%%%%%%%%%%%%

\subsection{Transformation to polar coordinates}

With the above results in place, we now transform the invariant $I(t)$ and the corresponding ladder operators to polar 
coordinates for calculational convenience. For this we invoke the transformation of coordinates of the form,
\begin{equation}
x=rcos\theta\,\,\,;\,\,\,y=rsin\theta~.
\label{eqn21}
\end{equation}
The canonical coordinates in polar representation takes the following form, 
\begin{eqnarray}
p_r&=&\dfrac{1}{2}\left(\dfrac{x_1}{r}{p_1}+{p_1}\dfrac{x_1}{r}+\dfrac{x_2}{r}{p_2}+p_2\dfrac{x_2}{r}\right)\nonumber\\
&=&\dfrac{x_1{p_1}+x_2{p_2}}{r}-\dfrac{i}{2r}\nonumber\\&=&-i\left({\partial}_r+\dfrac{1}{2r} \right)\\
p_{\theta}&=&(x_1{p_2}-x_2{p_1})=-i{\partial_{\theta}}.
\label{eqn22}
\end{eqnarray} 
The commutation relations between ($r$, $p_r$) and 
($\theta$, $p_\theta$) have the form
\begin{equation}
[r,p_r]=[\theta,p_{\theta}]=[x_1,p_1]=[x_2,p_2]=i.
\label{eqn23}
\end{equation}
The corresponding anticommutation relation can be found to be,
\begin{equation}
[r, p_r]_{+}=[x_1, p_1]_{+}+[x_2, p_2]_{+}
=2(x_1 p_1+p_2 x_2)
\label{eqn24}
\end{equation}
where $[A, B]_{+}=AB+BA$ represents anticommutator between operators $A$, $B$.

\noindent In order to transform the invariant $I(t)$ in polar coordinates, we need to have few other relations which are,
\begin{eqnarray}
({p_1}^2+{p_2}^2)&=&\left({p_r}^2+\dfrac{{p_{\theta}}^2}{r^2}-\dfrac{1}{4r^2}\right)\\
(p_1+i{p_2})&=&e^{i\theta}\left[p_r+\dfrac{i}{r}p_{\theta}+\dfrac{i}{2r} \right]      \\
(p_1-i{p_2})&=&e^{-i\theta}\left[p_r-\dfrac{i}{r}p_{\theta}+\dfrac{i}{2r} \right].
\label{eqn25}
\end{eqnarray}
Hence the invariant in polar coordinate system is given by,
\begin{eqnarray}
I(t)=\dfrac{\xi^2}{\rho^2}r^2+\left(\rho{p_r}-\dfrac{\dot{\rho}}{a}r\right)^2+\left({\dfrac{\rho{p_\theta}}{r}}\right)^2-\left({\dfrac{\rho\hbar}{2r}}\right)^2
\label{eqn26}
\end{eqnarray}
and the ladder operators in polar coordinate system have the following form,
\begin{eqnarray}
\hat{a}(t)&=&\dfrac{1}{2}\left[\left(\rho{p_r}-\dfrac{\dot{\rho}}{a(t)}r \right)-i\left(\dfrac{r}{\rho}+\dfrac{\rho{p_\theta}}{r}+\dfrac{\rho}{2r} \right)    \right]e^{-i\theta}\nonumber\\
{\hat{a}}^{\dagger}(t)&=&\dfrac{1}{2}e^{i\theta}\left[\left(\rho{p_r}-\dfrac{\dot{\rho}}{a(t)}r \right)+i\left(\dfrac{r}{\rho}+\dfrac{\rho{p_\theta}}{r}+\dfrac{\rho}{2r} \right) \right].
\label{eqn27}
\end{eqnarray}
Now we note from Eqn(s).(\ref{eqn26}, \ref{eqn27}) that both the invariant $I(t)$ 
and the ladder operators have the same form as those used in \cite{Dey} to study the undamped harmonic oscillator in 
NC space. The time-dependent coefficients involved in the 
present study however differ due to the damping present in our system. Thus, we 
can just borrow the expression of eigenfunction and the phase factors 
from \cite{Dey} for our present system.

%%%%%%%%%%%%%%%%%%%%%%%%%%%%%%%%%%%%%%%%%%%%%%%%%%%%%%%%%%%%%%%%%%%%%%%%%%%%%%%%%%%%%%%%%%%%%%%%%

%%%%%%%%%%%%%%%%%%%%%%%%%%%%%%%%%%%%%%%%%%%%%%%%%%%%%%%%%%%%%%%%%%%%%%%%%%%%%%%%%%%%%%%%%%%%%%%%%

\subsection{Eigenfunction and phase factor}
We depict the set of eigenstates of the invariant operator $I(t)$ as $\ket{n,l} $, following the convention in 
\cite{Dey}. Here, $n$ and $l$ are integers such that $n+l\geqslant0$. So we have the condition $l\geqslant-n$. 
Thus, if $l=-n+m$, then $m$ is a positive integer; and the corresponding eigenfunction in polar coordinate system has the following form (restoring $\hbar$), 
\begin{eqnarray}
\phi_{n,m-n}(r,\theta)&=&\braket{r,\theta|n,m-n}\\
&=&\lambda_{n}\dfrac{{(i\sqrt{\hbar}\rho)}^m}{\sqrt{m!}}r^{n-m}e^{i\theta(m-n)-\dfrac{a(t)-i\rho\dot{\rho}}{2a(t)
\hbar{\rho}^2}r^2}U\left(-m,1-m+n,\dfrac{r^2}{\hbar\rho^2} \right)
\label{eqn28}
\end{eqnarray}
where $\lambda_n$ is given by
\begin{eqnarray}
\lambda_n^2=\dfrac{1}{\pi{n!}{(\hbar\rho^2)}^{1+n}}~. 
\label{eqn28lam}
\end{eqnarray}
Here, $U\left(-m,1-m+n,\dfrac{r^2}{\hbar\rho^2} \right)$ is 
Tricomi's confluent hypergeometric function \cite{Arfken, uva} and the eigenfunction $\phi_{n,m-n}(r,\theta)$ satisfies the following 
orthonormality relation,
\begin{equation}
\int_0^{2\pi}d\theta\int_0^{\infty}rdr\phi^{*}_{n,m-n}(r,\theta)\phi_{n^{'},m^{'}-n^{'}}(r,\theta)=\delta_{nn^{'}}\delta_{mm^{'}}.
\label{eqn29}
\end{equation}
Again following \cite{Dey}, the expression of the phase factor $\Theta(t)$ is given by,  
\begin{equation}
\Theta_{\,n\,,\,l}(t)\,=\,(\,n\,+\,l\,)\,\int_0^t \left(c(T)-\dfrac{a(T)}{\rho^2(T)} \right)dT~.
\label{eqn30}
\end{equation}
For a given value of $l=-n+m$, it would be given by \cite{Dey},
\begin{equation}
\Theta_{\,n\,,\,m\,-\,n\,}(t)=m\int_0^t \left(c(T)-\dfrac{a(T)}{\rho^2(T)} \right)dT~.
\label{eqn31}
\end{equation}
We shall use this expression to compute the phase explicitly as a function of time for various physical cases in the subsequent discussion.

\noindent The eigenfunction of the Hamiltonian therefore reads (using Eqn(s).(\ref{eqnpsi}, \ref{eqn28}, \ref{eqn31}))
\begin{eqnarray}
\psi_{n,m-n}(r,\theta,t)&=&e^{i\Theta_{n, m-n}(t)}\phi_{n, m-n}(r,\theta)\nonumber\\
&=&\lambda_{n}\dfrac{{(i\sqrt{\hbar}\rho)}^m}{\sqrt{m!}}\exp{\left[im\int_0^t \left(c(T)-\dfrac{a(T)}{\rho^2(T)} \right)dT \right]}
\nonumber\\
&&\times~r^{n-m}e^{i\theta(m-n)-\dfrac{a(t)-i\rho\dot{\rho}}{2a(t)\hbar{\rho}^2}r^2}U\left(-m,1-m+n,\dfrac{r^2}{\hbar\rho^2} \right).
\label{eqn32}
\end{eqnarray}

%%%%%%%%%%%%%%%%%%%%%%%%%%%%%%%%%%%%%%%%%%%%%%%%%%%%%%%%%%%%%%%%%%%%%%%%%%%%%%%%%%%%%%%%%

%%%%%%%%%%%%%%%%%%%%%%%%%%%%%%%%%%%%%%%%%%%%%%%%%%%%%%%%%%%%%%%%%%%%%%%%%%%%%%%%%%%%%%%%%

\section{Solutions for the noncommutative damped oscillator}
In this paper we are primarily interested in damped oscillators in 
NC space. For this purpose we want to find the eigenfunctions of 
the corresponding Hamiltonian under various types of damping. The various 
kinds of damping are represented by various forms of the time dependent 
coefficients of the Hamiltonian, namely, $a(t)$, $b(t)$ and $c(t)$. 
However, the various forms must be constructed in such a way that they satisfy 
the non-linear EP equation given by Eqn.(\ref{eqn11}). 
The procedure of this construction of exact analytical solutions is based on the Chiellini integrability condition \cite{chill} and this formalism was followed in \cite{Dey}. We shall do the same in this paper.
So, for various forms of $a(t)$ and $b(t)$, we get the corresponding form of $\rho(t)$ 
using the EP equation together with the Chiellini integrability condition. In other words, the set of values of $a(t)$, $b(t)$ and $\rho(t)$ that we use must be a solution set of the EP equation consistent with the Chiellini integrability condition. In the subsequent discussion we shall proceed to obtain solutions of the EP equation for the damped NC oscillator.

%%%%%%%%%%%%%%%%%%%%%%%%%%%%%%%%%%%%%%%%%%%%%%%%%%%%%%%%%%%%%%%%%%%%%%%%%%%%%%%%%%%%%%%%%%

\subsection{Solution Set-I for Ermakov-Pinney equation : Exponentially \\ decaying solutions } 
\subsubsection{The Solution Set}
The simplest kind of solution set of EP equation under damping is the 
exponentially decaying set used in \cite{Dey}. The solution set is given by the following
relations, 
\begin{eqnarray}
a(t)=\sigma e^{-\vartheta{t}}\,\,\,,\,\,\,b(t)=\Delta e^{\vartheta{t}}\,\,\,,\,\,\rho(t)={\mu}e^{-\vartheta{t/2}}\,\,\,\,\,
\label{EPsoln1}
\end{eqnarray}
where $\sigma,\Delta$ and $\mu$ are constants. Here, $\vartheta$ is any 
positive real number. Substituting the expression of $a(t),b(t) \,$and$\, \rho(t)$ in the EP equation, we can easily verify the relation between these constants to be as follows, 
\begin{equation}
\mu^4=\dfrac{\xi^2{\sigma^2}}{\sigma\Delta-\dfrac{1}{4}\vartheta^2}~.
\label{EPreln1}
\end{equation}

%%%%%%%%%%%%%%%%%%%%%%%%%%%%%%%%%%%%%%%%%%%%%%%%%%%%%%%%%%%%%%%%%%%%%%%%%%%%%%%%%%%%%%%%%%

%%%%%%%%%%%%%%%%%%%%%%%%%%%%%%%%%%%%%%%%%%%%%%%%%%%%%%%%%%%%%%%%%%%%%%%%%%%%%%%%%%%%%%%%%%

\subsubsection{Study of the corresponding eigenfunctions}
We now write down the eigenfunctions of the Hamiltonian for the choosen set of 
time-dependent coefficients. For this endeavour we need to choose explicit 
forms of the damping factor $f(t)$ and angular frequency of the oscillator 
$\omega(t)$. The eigenfunction of the invariant $I(t)$ (which is given by 
Eqn.(\ref{eqn28})) takes on the following form for the solution set-I:
\begin{eqnarray}
\phi_{n,m-n}(r,\theta)=\lambda_{n}\dfrac{{(i{\mu}e^{-\vartheta{t/2}})}^m}{\sqrt{m!}}    r^{n-m}e^{i\theta(m-n)-\dfrac{2\sigma+i\mu^2\vartheta}{4\sigma\mu^2{e^{-\vartheta{t}}}}r^2}U\left(-m,1-m+n,\dfrac{r^2{e^{\vartheta{t}}}}{\mu^2} \right)
\label{eqn33}
\end{eqnarray}
where $\lambda_n$ is given by
\begin{eqnarray}
\lambda_n^{\,2}\,=\,\dfrac{1}{\pi\,n!\,[\mu^2\,exp\,(-\vartheta{t})]^{1+n}}~.
\label{eqn33lam}
\end{eqnarray}
In order to obtain explicit expressions of the phase factors for various cases of the damping factor, we choose both the 
functions $\omega(t)$ and $\eta(t)$ as follows.\\

%%%%%%%%%%%%%%%%%%%%%%%%%%%%%%%%%%%%%%%%%%%%%%%%%%%%%%%%%%%%%%%%%%%%%%%

%%%%%%%%%%%%%%%%%%%%%%%%%%%%%%%%%%%%%%%%%%%%%%%%%%%%%%%%%%%%%%%%%%%%%%%

\noindent{\bf $\langle A\rangle$ Solution Set-Ia}

\noindent Firstly, we choose the damping factor $f(t)=1$. Thus, in this case the damping in the system is due to the exponentially 
decaying frequency $\omega(t)$. For this purpose we set, 
\begin{eqnarray}
\eta(t)=0\,\,\Rightarrow\,\,f(t)=1\\
\omega(t)={\omega_0}\,exp(-\Gamma{t}/2)\,\,\,.
\label{eqn34}
\end{eqnarray}
Substituting the expressions for $a(t)$, $b(t)$, $\omega(t)$ and $f(t)$ in the 
Eqn(s).(\ref{3}, \ref{4}), we get the time dependent NC parameters as,
\begin{eqnarray}
\theta(t)=\dfrac{2}{M\omega_0}\,exp\,[\Gamma{t}/2] \sqrt{M\sigma\,exp(-\vartheta{t})-1}\label{eqn35}      \\
\Omega(t)=2\sqrt{\,M[\Delta\,exp\,(\vartheta{t})-M\omega_0^2\,exp\,(-\Gamma{t})]}. \label{eqn36}
\end{eqnarray}
It can be checked that in the limit $\Gamma\rightarrow0$, that is, for 
constant frequency, the expressions for $\theta(t)$ and $\Omega(t)$ reduce 
to those in \cite{Dey}. When $\vartheta=\Gamma$, then the solutions take the 
form,
\begin{eqnarray}
\theta(t)&=&\dfrac{2}{M\omega_0}\, \sqrt{M\sigma\,-e^{\Gamma t}}\label{eqn35b}      \\
\Omega(t)&=&2\sqrt{\,M[\Delta\,exp\,(\Gamma{t})-M\omega_0^2\,exp\,(-\Gamma{t})]}. \label{eqn36b}
\end{eqnarray}
Substituting these relations in the expression for $c(t)$ in Eqn.(\ref{eqn3}), 
we get an expression for the phase in a closed form as, 
\begin{equation}
c(t)=\sqrt{\dfrac{ \Delta\,exp\,(\Gamma{t})-M{\omega_0}^2\,exp\,(-\Gamma{t}) }{M}} 
\,+\,\omega_0\,exp\,(-\Gamma{t}/2)\sqrt{\,M\sigma\,exp\,(-\Gamma{t})-1 \,}.\label{eqn38}
\end{equation}
Substituting the expressions of $a(t)$, $\rho(t)$ and $c(t)$ in Eqn.(\ref{eqn31}), we get,
\begin{eqnarray}
\Theta_{\,n\,,\,l}(t)&=&\,(\,n\,+\,l\,)\,\dfrac{\omega_0}{2\sqrt{M\sigma}\Gamma}\ \left[log_{e}\dfrac{e^{{\Gamma}t}-2M\sigma-2\sqrt{M\sigma(M\sigma-e^{{\Gamma}t})}}{1-2M\sigma-2\sqrt{M\sigma(M\sigma-1)}}\right.\nonumber\\ 
&&\left. -{\Gamma}t-2\sqrt{M\sigma(M{\sigma}e^{-{2\Gamma}t}-e^{-{\Gamma}t})}+2\sqrt{M\sigma(M\sigma-1)}\ \right]\nonumber\\
&&+\dfrac{2(n+l)}{\Gamma}\,\left[\sqrt{\frac{\Delta}{M}e^{{\Gamma}t}-{\omega_0^2}e^{-{\Gamma}t}}-\sqrt{\dfrac{\Delta}{M}-{\omega_0^2}}\right.\nonumber\\
&&\left.+2i{\omega_0}\left\{e^{-{\Gamma}t/2} {_{2}F_{1}}\left(-\frac{1}{4},\frac{1}{2},\frac{3}{4},\frac{{\Delta}e^{{2\Gamma}t}}{M\omega_0^2}\right)- {_{2}F_{1}}\left(-\frac{1}{4},\frac{1}{2},\frac{3}{4},\frac{\Delta}{M\omega_0^2}\right)\right\}\right]
-\frac{\sigma}{\mu^2}(n+l)t  \nonumber\\
\label{eqn39}                      
\end{eqnarray}
where $_{2}F_{1}(a,b,c;t)$ is said to be the Gauss hypergeometric function. It is interesting to note that the solutions of the time dependent NC parameters enable us to get an exact analytic expression for the phase factor.
It is further interesting to observe that the phase has a complex part which indicates that the wave function decays with time.\\

%%%%%%%%%%%%%%%%%%%%%%%%%%%%%%%%%%%%%%%%%%%%%%%%%%%%%%%%%%%%%%%%%%%%%%%%%%%%%%

%%%%%%%%%%%%%%%%%%%%%%%%%%%%%%%%%%%%%%%%%%%%%%%%%%%%%%%%%%%%%%%%%%%%%%%%%%%%%%

\noindent {\bf $\langle B\rangle$ Solution Set-Ib}

\noindent Here the oscillator is damped due to the damping factor $f(t)$ and the frequency $\omega(t)$ 
is a constant. This situation can be depicted by the following relations, 
\begin{eqnarray}
f(t)= exp\,(-\Gamma{t})~;~\omega(t)={\omega_0}.
\label{10x}
\end{eqnarray}
Substituting these relations in Eqn(s).(\ref{3}, \ref{4}), we get the time dependent NC parameters 
as,
\begin{eqnarray}
\theta(t)&=&\dfrac{2}{M\omega_0}\sqrt{M{\sigma}\,exp\,(-\vartheta{t})-exp\,(-\Gamma{t})}\,\,e^{-\Gamma t/2}\label{eqn40}\\
\Omega(t)&=&2e^{\Gamma{t}}\sqrt{M\,[\Delta\,exp\,(\vartheta-\Gamma)t-M{\omega_0}^2]}.
\label{eqn41}
\end{eqnarray}
It can be checked that in the limit $\Gamma\rightarrow0$, that is, for 
constant frequency, the expressions for $\theta(t)$ and $\Omega(t)$ reduce 
to those in \cite{Dey}. When $\vartheta=\Gamma$, then the solutions take the 
form,
\begin{eqnarray}
\theta(t)&=&\dfrac{2}{M\omega_0}\sqrt{M{\sigma}\,- 1}\,\,e^{-\Gamma t}\label{eqn40b}\\
\Omega(t)&=&2e^{\Gamma{t}}\sqrt{M\,[\Delta\,-M{\omega_0}^2]}.
\label{eqn41b}
\end{eqnarray}
Substituting these relations in the expression for $c(t)$ in Eqn.(\ref{eqn3}), we 
get,
\begin{equation}
c(t)= \sqrt{\dfrac{ \Delta\,-M{\omega_0}^2\, }{M}} \,+\,\omega_0\sqrt{M\sigma-1}\,=\,constant ~. \label{eqn43}
\end{equation}
Substituting the expressions of $a(t)\,,\rho(t)$\, and $c(t)$\, in Eqn.(\ref{eqn31})\,,\,we get an expression for the phase in a closed form as,
\begin{align}
\Theta_{\,n\,,\,l}(t)\,=&(\,n\,+\,l\,)\left[-\frac{\sigma}{\mu^2}\,+\,\sqrt{\frac{ \Delta\,-M{\omega_0}^2\,}{M} } \,+\omega_0\sqrt{M\sigma-1}  \right]\,t ~.\label{eqn44}
\end{align}
Once again we are able to obtain an exact expression for the phase, in this case varying linearly with time.
It is important to note that the reality of the phase in this case depends crucially on the parameters $\Delta$, $M$, $\sigma$, $\omega_0$. The phase $\Theta_{n,l}$ is real if $\Delta-M\omega_{0}^2 \geq 0$ and $M\sigma \geq 1$, else it is complex.\\

%%%%%%%%%%%%%%%%%%%%%%%%%%%%%%%%%%%%%%%%%%%%%%%%%%%%%%%%%%%%%%%%%%%%%%%%%%%%%%%%%%%%

%%%%%%%%%%%%%%%%%%%%%%%%%%%%%%%%%%%%%%%%%%%%%%%%%%%%%%%%%%%%%%%%%%%%%%%%%%%%%%%%%%%%

\noindent {\bf $\langle C\rangle$ Solution Set-Ic}

\noindent Here the oscillator is damped due to the damping factor $f(t)$ and the time-dependent 
frequency $\omega(t)$; both of which are exponentially decaying. Thus, we set, 
\begin{eqnarray}
f(t)=exp\,(-\Gamma{t})~;~\omega(t)={\omega_0}\,exp\,(-\Gamma{t}/2).
\label{eqn45}
\end{eqnarray}
Substituting these relations in Eqn.(s)(\ref{3}, \ref{4}), we get the time dependent NC 
parameters to be, 
\begin{eqnarray}
\theta(t)&=&\dfrac{2}{M\omega_0}\sqrt{(M{\sigma}e^{-(\vartheta-\Gamma)t}\,- 1)}e^{-\Gamma{t}/2}\label{eqn46}\\
\Omega(t)&=&2\sqrt{M\,[\Delta\,exp(\vartheta{t})-M{\omega_0}^2\, \,]}\,\,e^{\Gamma t/2}.\label{eqn47}
\end{eqnarray}
It can be checked that in the limit $\Gamma\rightarrow0$, that is, for 
constant frequency, the expressions for $\theta(t)$ and $\Omega(t)$ reduce 
to those in \cite{Dey}. When $\vartheta=\Gamma$, then the solutions take the 
form,
\begin{eqnarray}
\theta(t)&=&\dfrac{2}{M\omega_0}\sqrt{(M{\sigma}\,- 1)}e^{-\Gamma{t}/2}\,\,\label{eqn46}\\
\Omega(t)&=&2\sqrt{M\,[\Delta\,exp(\Gamma{t})-M{\omega_0}^2\, \,]}\,\,e^{\Gamma t/2}.\label{eqn47}
\end{eqnarray}
Substituting these relations in the expression for $c(t)$ in Eqn.(\ref{eqn3}), we 
get,
\begin{align}
c(t)= \sqrt{\dfrac{ \Delta\,-M{\omega_0}^2\,\exp\left[-\Gamma{t}\right] }{M}} 
 +  {\omega_0}\,e^{-\Gamma t/2}\sqrt{M{\sigma}\,-1}~.\label{eqn49}
\end{align}
Substituting the expressions of $a(t)$, $\rho(t)$ and $c(t)$\, in Eqn.(\ref{eqn31}), we obtain an expression for the phase in a closed form as,
\begin{eqnarray}
\Theta_{\,n\,,\,l}(t)\,&=&\,\dfrac{(\,n+l\,)}{\Gamma\,\sqrt{M}}\,\left[\,\sqrt{\Delta}\,\Gamma\,t\,+\,2\sqrt{\Delta-M\omega_0^2}\,-2\sqrt{\Delta-M\omega_0^2{\exp\,(-\Gamma{t})}}\right.\nonumber\\
&&\left.+\,2\sqrt{\Delta}\,log\,\left(\frac{\Delta+\sqrt{\Delta[\Delta-M\omega_0^2\,\exp\,(-\Gamma{t}) ]}}{\Delta+\sqrt{\Delta[\Delta-M\omega_0^2\,]}}\right)     \right]\nonumber \\
&&-\,(\,n+l\,)\left[\dfrac{\sigma\,t}{\mu^2}\,+\,\dfrac{2}{\Gamma}\,\omega_0\,\left(e^{-\Gamma t/2}-1\right)\sqrt{\,M\sigma-1} \right].\label{eqn50}
\end{eqnarray}

%%%%%%%%%%%%%%%%%%%%%%%%%%%%%%%%%%%%%%%%%%%%%%%%%%%%%%%%%%%%%%%%%%%%%%%%%%%%%%%%%%%%%%%%%%%%%%%%%%%%%

%%%%%%%%%%%%%%%%%%%%%%%%%%%%%%%%%%%%%%%%%%%%%%%%%%%%%%%%%%%%%%%%%%%%%%%%%%%%%%%%%%%%%%%%%%%%%%%%%%%%%

\subsection{Solution Set-II for Ermakov-Pinney equation: Rationally decaying solutions}
\subsubsection{The Solution Set}
We now consider rationally decaying solutions of the EP equation similar to that used in~\cite{Dey} which is of the form,
\begin{eqnarray}
&a(t)=\dfrac{\sigma\,\left(1+\dfrac{2}{k}\right)^{\,(k+2)/k}}{(\Gamma{t}+\chi)^{\,(k+2)/k}}\nonumber \\ \nonumber\\
&b(t)=\dfrac{\Delta\,\left(\dfrac{k}{k+2} \right)^{(2-k)/k} }{(\Gamma{t}+\chi)^{\,(k-2)/k}}     \,\,\,\Rightarrow\, \,\,\,\dfrac{\Delta\,\left(1+\dfrac{2}{k}\right)^{\,(k-2)/k} }{(\Gamma{t}+\chi)^{\,(k-2)/k}}\nonumber  \\ \nonumber\\
&\rho(t)=\dfrac{\mu\left(1+\dfrac{2}{k}\right)^{1/k} }{(\Gamma{t}+\chi)^{1/k}}
\label{EPsoln2}
\end{eqnarray}
where $\sigma$, $\Delta$, $\mu$, $\Gamma$ and $\chi$ are constants such that $(\Gamma{t}+\chi)~\neq~0$, and $k$ is an integer. Substituting the expressions of $a(t)$, $b(t)$, and $\rho(t)$ in the EP equation, we can easily verify the relation between these constants to be as follows, 
\begin{equation}
\Gamma^2\mu=(k+2)^2\,(\sigma\Delta\mu-\frac{\xi^2\sigma^2}{\mu^3}).
\label{EPreln2}
\end{equation}

%%%%%%%%%%%%%%%%%%%%%%%%%%%%%%%%%%%%%%%%%%%%%%%%%%%%%%%%%%%%%%%%%%%%%%%%%%%%%%%%%%

%%%%%%%%%%%%%%%%%%%%%%%%%%%%%%%%%%%%%%%%%%%%%%%%%%%%%%%%%%%%%%%%%%%%%%%%%%%%%%%%%%

\subsubsection{Study of the corresponding eigenfunctions}
The eigenfunction of the invariant operator $I(t)$ (given by 
Eqn.(\ref{eqn28})) for this solution Set-II is given by,
\begin{eqnarray}
\phi_{n\,,\,m-n}(r,\theta)=\lambda_{n}\,\dfrac{{(i\mu)}^{\,m}}{\sqrt{m!}}\left[\dfrac{k+2}{k(\Gamma{t}+\chi)}\right]^{m/k}    r^{n-m}e^{i\theta(m-n)-\dfrac{[\sigma\,(k+2)\,+\,i\mu^2\Gamma]\,\,(\Gamma{t}+\chi)^{2/k}\,\,\,k^{2/k} }{2\sigma\,(k+2)^{\,(k+2)/k}\mu^2}r^2}\nonumber \\
\times\,\,\,U\left(-m,1-m+n,\,\dfrac{r^2[k(\Gamma{t}+\chi)]^{2/k}}{\mu^2\left(k+2\right)^{2/k}}\,\right)
\label{eqn51}
\end{eqnarray}
where $\lambda_n$ is given by 
\begin{eqnarray}
\lambda_n^{\,2}=\dfrac{1}{\pi\,n!\mu^{2n+2}}\left[\dfrac{k(\Gamma{t}+\chi)}{k+2}\right]^{2(1+n)/k}.
\label{eqn51lam}
\end{eqnarray}
In order to get the eigenfunction of the Hamiltonian $H(t)$, we need to calculate the associated phase factor. Once again for this we need to fix up the forms of the damping factor $f(t)$ and angular frequency $\omega(t)$ of the oscillator. In order to 
explore the solution of $H(t)$ for rationally decaying coefficients, we choose a rationally decaying form for $\omega(t)$ 
and set $f(t)=1$. Thus, we have the following relations, 
\begin{eqnarray}
\eta(t)=0\,\,\Rightarrow\,\,f(t)=1\\
\omega(t)=\dfrac{\omega_0}{(\Gamma\,t+\chi)}~.
\end{eqnarray}
Substituting these relations in Eqns.(\ref{3}, \ref{4}), we get the time dependent NC parameters as, 
\begin{eqnarray}
\theta(t)&=& \dfrac{2\,(\Gamma\,t+\chi)}{M\,\omega_0}\,\sqrt{M\sigma\,\left[\dfrac{(k+2)}{k\,(\Gamma{t}\,+\,\chi)}\right]^{(k+2)/k}\,-\,1}     \label{eqn52}                   \\ \nonumber \\
\Omega(t)&=& \,2\,\sqrt{M\Delta\,\left[\dfrac{k+2}{k(\Gamma{t}+\chi)}\right]^{\,(k-2)/k}\,-\,\dfrac{M^{\,2}\omega_0^{\,2}}{(\Gamma\,t+\chi)^2}}~.\label{eqn53}
\end{eqnarray}
We now consider $k=2$. This enables us to integrate the expression for the phase factor (given by Eqn.(\ref{eqn31})).
The simplified forms of $a(t)$, $b(t)$ and $\rho(t)$ for $k=2$ read,
\begin{eqnarray}
a(t)=\dfrac{4\sigma}{(\Gamma{t}+\chi)^{\,2}}\,\,,\,\,b(t)\,=\,\Delta\,\,,\,\,\rho(t)=\left[\dfrac{2\mu^{\,2}}{\Gamma{t}+\chi}\right]^{1/2}.
\end{eqnarray}
Substituting these relations in the expression for $c(t)$ in Eqn.(\ref{eqn3}) gives, 
\begin{equation}
c(t)\,=\,\dfrac{\omega_0}{(\Gamma\,t+\chi)}\,\sqrt{\dfrac{4\sigma\,M}{(\Gamma{t}+\chi)^2}\,-\,1}\,+\,\sqrt{\dfrac{\Delta}{M}\,-\,\dfrac{\omega_0^{\,2}}{(\Gamma\,t+\chi)^{\,2}}}\,\,\,\,.
\end{equation} 
Substituting these expressions for $a(t)$, $\rho(t)$ and $c(t)$ 
for $k=2$ in Eqn.(\ref{eqn31}), we get the following expression 
for the phase factor in a closed form as,
\begin{eqnarray}
\Theta_{\,n, l\,}(t)&=&\dfrac{(n+l)}{\Gamma}\,\left[\omega_0\,\,tan^{\,-1}\left(\dfrac{\omega_0}{\sqrt{\frac{\Delta}{M}{(\Gamma\,t+\chi)^2}-\omega_0^2}}\right)+\sqrt{\dfrac{\Delta\,{(\Gamma\,t+\chi)^2}}{M}-\omega_0^2}\,-\,\frac{2\sigma}{\mu^2}\,log_{e}\,\frac{(\chi+\Gamma\,t)}{\chi}\right.\nonumber\\
&&\left.-{\sqrt{\frac{\Delta}{M}{\chi^2}-\omega_0^2}}-\omega_0\,\,tan^{\,-1}\left(\dfrac{\omega_0}{\sqrt{\frac{\Delta}{M}{\chi^2}-\omega_0^2}}\right)\right] \nonumber\\
&& +\dfrac{\omega_0(n+l)}{\Gamma}\left[\dfrac{\sqrt{4\,\sigma\,M-\chi^2}}{\chi}-\dfrac{\sqrt{4\,\sigma\,M-(\chi+\Gamma\,t)^2}}{(\chi+\Gamma\,t)}
\right.\nonumber\\
&&\left. +ilog_{e}\dfrac{(\chi+\Gamma\,t)+{\sqrt{(\chi+\Gamma\,t)^2-4\,\sigma\,M}}}{\chi+\sqrt{\chi^2-4\,\sigma\,M}} \right].
\label{eqn55}
\end{eqnarray}
We can now get the eigenfunction of this rationally decaying damped system using 
Eqn.(\ref{eqnpsi}).

%%%%%%%%%%%%%%%%%%%%%%%%%%%%%%%%%%%%%%%%%%%%%%%%%%%%%%%%%%%%%%%%%%%%%%%%%%%%%%%%%%%%%%%%

%%%%%%%%%%%%%%%%%%%%%%%%%%%%%%%%%%%%%%%%%%%%%%%%%%%%%%%%%%%%%%%%%%%%%%%%%%%%%%%%%%%%%%%%

\subsection{Solution Set-III for Ermakov-Pinney equation: Elementary Solution}
\subsubsection{The Solution Set}
We now propose a simple method of obtaining a solution of the EP equation.
The method is as follows. Choosing $\rho(t)$ to be any arbitrary time dependent function and taking it's time derivative as proportional to $a(t)$, that is,
$a(t)=constant \times\dot{\rho}$ and setting 
$b(t)=constant \times \dfrac{a}{\rho^4}$, we observe that these would always satisfy the EP equation along with a certain constraint relation among the constants.

\noindent Here we consider a simple solution which is a special case of the above  solution for the EP equation. We call this the elementary solution which reads,
\begin{eqnarray}
a(t)={\sigma}\,\,\,\,,\,\,\,b(t)=\dfrac{{\Delta}}{{(\Gamma\,t\,+\,\chi)^4}}\,\,\,,\,\,\rho(t)=\mu(\Gamma{t}\,+\,\chi)
\label{eqn57}
\end{eqnarray}
where $\Gamma$, $\chi$, $\mu$, $\sigma$ and $\Delta$ are constants. The above solution set satisfy the EP equation with the following constraint relation,
\begin{equation}
\Delta\mu^4=\xi^2\sigma\,\,.
\end{equation}

%%%%%%%%%%%%%%%%%%%%%%%%%%%%%%%%%%%%%%%%%%%%%%%%%%%%%%%%%%%%%%%%%%%%%%%%%%%%%%%

%%%%%%%%%%%%%%%%%%%%%%%%%%%%%%%%%%%%%%%%%%%%%%%%%%%%%%%%%%%%%%%%%%%%%%%%%%%%%%%

\subsubsection{Study of the corresponding eigenfunctions}
The eigenfunctions of the invariant operator $I(t)$ for this solution set is given by,
\begin{eqnarray}
\phi_{n,m-n}(r,\theta)&=&\lambda_{n}\dfrac{{[i\mu(\Gamma\,t+\chi)]}^m}{\sqrt{m!}}r^{n-m}e^{i\theta(m-n)-\dfrac{\sigma-i\mu^2\Gamma(\Gamma\,t+\chi)}{2\sigma\mu^2(\Gamma\,t+\chi)^2}r^2}\nonumber \\
&&\times\,\,U\left(-m,1-m+n,\dfrac{r^2}{\mu^2(\Gamma\,t+\chi)^2} \right)
\label{eqn58}
\end{eqnarray}
where $\lambda_n$ is given by
\begin{eqnarray}
\lambda_n^2=\dfrac{1}{\pi{n!}{\left[\mu(\Gamma\,t+\chi) \right]}^{2+2n}}~. 
\label{eqn58lam}
\end{eqnarray}
In order to get an eigenfunction of the Hamiltonian, we calculate the phase factor for a particular case of the damped harmonic oscillator where the angular frequency $\omega(t)$ is rationally decaying and the damping factor $f(t)$=1. Thus, we set, 
\begin{eqnarray}
\eta(t)=0\,\,\Rightarrow\,\,f(t)=1\\
\omega(t)=\dfrac{\omega_0}{(\Gamma\,t+\chi)}
\end{eqnarray}
where $\Gamma$ and $\chi$ are real constants.
Substituting these relations in Eqns.(\ref{3}, \ref{4}), we get the time dependent NC parameters as, 
\begin{eqnarray}
\theta(t)=\dfrac{2\,(\Gamma\,t+\chi)}{\omega_0\,M}\,\sqrt{M\,\sigma-1}\label{eqn59}\\
\Omega(t)=2\sqrt{\dfrac{M\Delta}{(\Gamma\,t+\chi)^4}-\dfrac{M^2\,\omega_0^2}{(\Gamma\,t+\chi)^{2}}}~.\label{eqn60}
\end{eqnarray}
Substituting these relations 
in the expression for $c(t)$ in Eqn.(\ref{eqn3}), we get,  
\begin{align}
c(t)=\sqrt{\dfrac{\Delta}{M(\Gamma\,t+\chi)^4}-\dfrac{\omega_0^2}{(\Gamma\,t+\chi)^2}}\,+\,\dfrac{\omega_0}{(\Gamma\,t+\chi)}\sqrt{M\,\sigma-1}.\label{eqn61}
\end{align}
Substituting these expressions of $a(t)$, $\rho(t)$ and $c(t)$ in Eqn.(\ref{eqn31}), we obtain an expression for the phase factor in a closed form as, 
\begin{align}
\Theta_{\,n\,,\,l}(t)&=\,(n+l\,)\left[\omega_0\dfrac{\sqrt{M\,\sigma-1}}{\Gamma}\log\,\dfrac{(\Gamma\,t+\chi)}{\chi}-\frac{{\sigma}t}{\mu^2\chi(\Gamma\,t+\chi)} \right] +\,\dfrac{(\,n+l\,)}{\Gamma}\left[\sqrt{\dfrac{\Delta}{M\chi^2}-\omega_0^2}\right.\nonumber\\
&\left.-\sqrt{\dfrac{\Delta}{M(\Gamma\,t+\chi)^2}-\omega_0^2}
+\omega_0\left\{\tan^{-1}\left(\dfrac{\omega_0\chi}{\sqrt{\dfrac{\Delta}{M}-\chi^2\omega_0^2}}\right)-\tan^{-1}\left(\dfrac{\omega_0(\Gamma\,t+\chi)}{\sqrt{\dfrac{\Delta}{M}-\omega_0^2(\Gamma\,t+\chi)^2}}\right)\right\}\,\, \right].
\label{eqn62}
\end{align} 
We can now get the eigenfunction of this system by using Eqn.(\ref{eqnpsi}).

%%%%%%%%%%%%%%%%%%%%%%%%%%%%%%%%%%%%%%%%%%%%%%%%%%%%%%%%%%%%%%%%%%%%%%%%%%%%%%%%%%%%%%%%%%%%%

%%%%%%%%%%%%%%%%%%%%%%%%%%%%%%%%%%%%%%%%%%%%%%%%%%%%%%%%%%%%%%%%%%%%%%%%%%%%%%%%%%%%%%%%

\section{Expectation Values}
In this section, we intend to calculate the expectation value of energy. For 
this we need to calculate the expectation value of the Hamiltonian $H(t)$ in 
it's own eigenstates. 
The expectation value $\langle H \rangle$ is given by (using Eqn.(\ref{eqn2})), 
\begin{equation}
\langle H\rangle = \dfrac{a(t)}{2}(\langle{p_1}^2\rangle+\langle{p_2}^2\rangle)+\dfrac{b(t)}{2}(\langle{x_1}^2\rangle+\langle{x_2}^2\rangle)+c(t)(\langle{p_1}{x_2}\rangle-\langle{p_2}{x_1}\rangle)\,\,\,.\label{eqn63}
\end{equation}
To calculate this we need to get the expectation value of the individual canonical 
operators. To set up our notation we denote the eigenstates of the Hamiltonian $H(t)$  
by $|n,l\rangle_H$. 

%%%%%%%%%%%%%%%%%%%%%%%%%%%%%%%%%%%%%%%%%%%%%%%%%%%%%%%%%%%%%%

%%%%%%%%%%%%%%%%%%%%%%%%%%%%%%%%%%%%%%%%%%%%%%%%%%%%%%%%%%%%%%

\subsection{Matrix elements of the coordinate operators raised to arbitrary finite powers}
\noindent We start by calculating the matrix element of an arbitrary power of $x$, $_{H}\langle n,l|x^k|n,l\rangle_{H}$, which is given by
\begin{eqnarray} 
_{H}\langle n,m-n|x^k|n,m'-n\rangle_{H}&=&\int r dr d\theta
~_{H}\langle n,m-n |r,\theta\rangle\langle r,\theta|r^k cos^k\theta|n,m'-n\rangle_{H}\nonumber\\
&=&\dfrac{1}{2^k}e^{i(\Theta_{n,m'-n}-\Theta_{n,m-n})}\int r^{k+1} dr d\theta~(e^{i\theta}+e^{-i\theta})^k \nonumber\\
&&~~~~~~~~~~~~~~~~~~~~~~~~~~~~~~~~\times\phi^*_{n,m-n}(r,\theta)\phi_{n,m'-n}(r,\theta)\nonumber\\ 
\label{eqn64}
\end{eqnarray} 
where we have used the relations, 
$|n,l\rangle_{H}~=~e^{i\Theta_{n,l}}|n,l\rangle$ where $|n,l\rangle_H$ and 
$|n,l\rangle$ are eigenstates of the Hamiltonian $H(t)$ and Lewis invariant $I(t)$ 
respectively. We have also used the relation $\langle r,\theta|n,m'-n\rangle$~=~$\phi_{n,m'-n}(r,\theta)$, with $\phi$ being the eigenfunction of $I(t)$.
Now, Eqn.(\ref{eqn64}) can be rewritten as,
\begin{eqnarray}
_{H}\langle n,m-n|x^k|n,m'-n\rangle_{H}&=&\dfrac{\pi}{2^{k-1}} \sum_{r=0}^{k} {^{k}C_r}\delta_{m',m+2r-k}  A(n,m,m+2r-k)\nonumber\\
&\times& \int_0^\infty r~dr~r^{2(n-m-r+k)}
e^{\dfrac{-r^2}{\hbar \rho^2}} \nonumber\\
&\times& U\left(-m,1-m+n,\dfrac{r^2}{\hbar\rho^2} \right)\nonumber\\
&\times& U\left(-m-2r+k,1-m-2r+k+n,\dfrac{r^2}{\hbar\rho^2} \right)
\label{eqn67}
\end{eqnarray}
where $A(n,m,m+2r-k)=e^{i(\Theta_{n,m-n+2r-k}-\Theta_{n,m-n})}\lambda_n^2\dfrac{(-i\hbar^{1/2}\rho)^{m}(i\hbar^{1/2}\rho)^{m+2r-k}}{\sqrt{m!(m+2r-k)!}}$~. 

\noindent Now defining $w=-\dfrac{r^2}{\hbar \rho^2}$, we have, 
\begin{eqnarray}
_{H}\langle n,m-n|x^k|n,m'-n\rangle_{H}&=&\sum_{r=0}^{k}\frac{\pi}{2^k}e^{i(\Theta_{n,m-n+2r-k}-\Theta_{n,m-n})}(-1)^{k+r}i^{-k}~{^{k}C_r}\delta_{m',m+2r-k}\nonumber\\
&\times& \lambda_n^2(\hbar^{1/2} \rho)^{2n+k+2}\sqrt{m!(m+2r-k)!}\nonumber\\
&\times& \int_0^\infty dw w^{n-m-r+k}e^{-w} L_m^{(n-m)}(w) L_{m+2r-k}^{(n-m-2r+k)}(w)
\label{eqn68}
\end{eqnarray}
where we have used the following result on special functions \cite{Arfken, uva},
\begin{eqnarray}
L_n^{(\zeta)}(w)=\frac{(-1)^n}{n!}U(-n,\zeta+1,w) \label{eqn69}
% L_n^{(\zeta)}(x)=L_n^{(\zeta+1)}(x)-L_{n-1}^{(\zeta+1)}(x).
\label{eqn70}
\end{eqnarray}
where $L_n^{(\zeta)}(w)$ are associated Laguerre polynomials.

\noindent Now, we get using the relation for phase given in \cite{Dey},
\begin{eqnarray}
\Theta_{n,l}=(n+l)\int^t \left[c(\tau)-\frac{a(\tau)}{\rho^2 (\tau)}\right]d\tau
\label{eqn73}
\end{eqnarray} 
the following relation,
\begin{eqnarray}
e^{i(\Theta_{n,m-n+2r-k}-\Theta_{n,m-n})}&=&e^{i[\{(n+m-n+2r-k)-(n+m-n)\}\int^t (c(\tau)-\frac{a(\tau)}{\rho^2 (\tau)})d\tau]} \nonumber\\
&=& e^{i(0+2r-k)\int^t (c(\tau)-\frac{a(\tau)}{\rho^2 (\tau)})d\tau}\nonumber\\
&=&e^{i\Theta_{0,2r-k}}.
\label{eqn74}
\end{eqnarray}
So, we finally get the following relation for the matrix element of $x^k$,
\begin{eqnarray}
_{H}\langle n,m-n|x^k|n,m'-n\rangle_{H}&=&\sum_{r=0}^{k}\frac{\pi}{2^k}e^{i\Theta_{0,2r-k}}(-1)^{k+r}i^{-k}~{^{k}C_r}\delta_{m',m+2r-k}\nonumber\\
&\times& \lambda_n^2(\hbar^{1/2} \rho)^{2n+k+2}\sqrt{m!(m+2r-k)!}\nonumber\\
&\times& \int_0^\infty dw~w^{n-m-r+k}e^{-w} L_m^{(n-m)}(w) L_{m+2r-k}^{(n-m-2r+k)}(w).
\label{eqn75}
\end{eqnarray}
This is a new result in this paper and can be used to obtain the matrix element or expectation value of any power of $x$.
For the sake of completeness, we also write down the matrix element of $x^k$ in the eigenstates of the Lewis invariant $I(t)$, which reads
\begin{eqnarray}
\langle n,m-n|x^k|n,m'-n\rangle&=&\sum_{r=0}^{k}\frac{\pi}{2^k}(-1)^{k+r}i^{-k}~{^{k}C_r}\delta_{m',m+2r-k}\nonumber\\
&\times& \lambda_n^2(\hbar^{1/2} \rho)^{2n+k+2}\sqrt{m!(m+2r-k)!}\nonumber\\
&\times& \int_0^\infty dw~w^{n-m-r+k}e^{-w} L_m^{(n-m)}(w) L_{m+2r-k}^{(n-m-2r+k)}(w).
\label{eqn75I}
\end{eqnarray}
Note that the phase factor does not appear in the above result.

\noindent Now, we proceed to evaluate the matrix element 
$_{H}\langle n,m-n|x|n,m'-n\rangle_{H}$ using the expression  
obtained in Eqn.(\ref{eqn75}). This reads
\begin{eqnarray}
_{H}\langle n,m-n|x|n,m'-n\rangle_{H}=
_{H}\langle n,m-n|x^k\mid_{\,k=1;\,r=0}|n,m'-n\rangle_{H}\nonumber\\
+ _{H}\langle n,m-n|x^k\mid_{\,k=1;\,r=1}|n,m'-n\rangle_{H}.
\label{eqn76}
\end{eqnarray}
Evaluating the above matrix elements give,
\begin{eqnarray}
_{H}\langle n,m-n|x^k\mid_{\,k=1;\,r=0}|n,m'-n\rangle_{H}= -\frac{i}{2}
{(\rho\hbar^{1/2})}\sqrt{m}e^{-i\Theta_0,1}\delta_{m,m'+1}
\label{eqn77}
\end{eqnarray}
\begin{eqnarray}
_{H}\langle n,m-n|x^k\mid_{\,k=1;\,r=1}|n,m'-n\rangle_{H}=\frac{i}{2}
{(\rho\hbar^{1/2})}\sqrt{m'}e^{i\Theta_0,1}\delta_{m',m+1}.
\label{eqn78}
\end{eqnarray}
In order to obtain Eqn(s).(\ref{eqn77}, \ref{eqn78}), we  
used the following relations involving the associated Laguerre polynomials,
\begin{eqnarray}
L_n^{(\zeta)}(w)=L_n^{(\zeta+1)}(w)-L_{n-1}^{(\zeta+1)}(w) \nonumber\\
\int_0^\infty dw~w^{\zeta} e^{-w}L_n^{(\zeta)} (w)L_m^\zeta (w) =\frac{(n+\zeta)!}{n!}\delta_{n,m}~.
\label{eqn80}
\end{eqnarray}
Combining Eqn(s).(\ref{eqn77}, \ref{eqn78}), we get the following expression,
\begin{eqnarray}
_{H}\langle n,m-n|x|n,m'-n\rangle_{H}~=~\frac{i}{2}
{(\rho\hbar^{1/2})}[\sqrt{m'}e^{i\Theta_{0,1}}\delta_{m',m+1}
-\sqrt{m}e^{-i\Theta_{0,1}}\delta_{m,m'+1}].
\label{eqn79}
\end{eqnarray}
Next, we evaluate,
\begin{eqnarray}
_{H}\langle n,m-n|x^2|n,m'-n\rangle_{H}= _{H}\langle n,m-n|x^k\mid_{\,k=2;\,r=0}|n,m'-n\rangle_{H}\nonumber\\
+_{H}\langle n,m-n|x^k\mid_{\,k=2;\,r=1}|n,m'-n\rangle_{H}+ _{H}\langle n,m-n|x^k\mid_{\,k=2;\,r=2}|n,m'-n\rangle_{H}. 
\label{eqn81}
\end{eqnarray}
Evaluation of the above matrix elements yield, 
\begin{eqnarray}
_{H}\langle n,m-n|x^k\mid_{\,k=2;\,r=0}|n,m'-n\rangle_{H}&=& 
-\frac{1}{4}{(\hbar\rho^2)}e^{-i\Theta_{0,2}}\delta_{m',m-2}\sqrt{m(m-1)}\nonumber\\
_{H}\langle n,m-n|x^k\mid_{\,k=2;\,r=1}|n,m'-n\rangle_{H}&=&~\frac{1}{2}{(\hbar\rho^2)}e^{-i\Theta_{0,0}}\delta_{m,m'}(m+n+1)\nonumber\\
_{H}\langle n,m-n|x^k\mid_{\,k=2;\,r=2}|n,m'-n\rangle_{H}&=& 
-\frac{1}{4}{(\hbar\rho^2)}e^{i\Theta_{0,2}}\delta_{m',m+2}\sqrt{(m+2)(m+1)}.
\label{eqn82}
\end{eqnarray}
In order to calculate the above expressions, apart from the relations between 
special functions given by Eqn.(\ref{eqn80}), we need the following relation,
\begin{eqnarray}
\int_{0}^{\infty} dw~w^{k+p}e^{-w}L_n^k(w)L_n^k(w)~=~\frac{(n+k)!}{n!}\times(2n+k+1)^p ~.
\label{eqn83}
\end{eqnarray}
So we have,
\begin{eqnarray}
_{H}\langle n,m-n|x^2|n,m'-n\rangle_{H}&=&\frac{(\hbar\rho^2)}{2}\delta_{m,m'}(m+n+1)\nonumber\\
&&-\frac{(\hbar\rho^2)}{4}\left[e^{-i\Theta_{0,2}}\delta_{m',m-2}\sqrt{m(m-1)}\right.\nonumber\\
&& \left. +e^{i\Theta_{0,2}}\delta_{m',m+2}\sqrt{(m+2)(m+1)}\right].
\label{eqn84}
\end{eqnarray}
It is to be noted that the matrix elements for $x$ and $x^2$ in the 
eigenstates of the Hamiltonian [given by Eqn(s).(\ref{eqn79}, \ref{eqn84}) 
respectively], matches exactly with the corresponding expression 
given in \cite{Dey}, although the result 
quoted in \cite{Dey} is in the eigenstate of the invariant $I(t)$.

\noindent The matrix element of $y^k$ in the 
eigenstates of the Hamiltonian can be obtained similarly, and reads,
\begin{eqnarray}
_{H}\langle n,m-n|y^k|n,m'-n\rangle_{H}&=&\sum_{r=0}^{k}\frac{\pi}{2^k}e^{i\Theta_{0,2r-k}}~{^{k}C_r}\delta_{m',m+2r-k}\nonumber\\
&\times& \lambda_n^2(\hbar^{1/2} \rho)^{2n+k+2}\sqrt{m!(m+2r-k)!}\nonumber\\
&\times& \int_0^\infty dw~w^{n-m-r+k}e^{-w} L_m^{(n-m)}(w) L_{m+2r-k}^{(n-m-2r+k)}(w).
\label{eqn85}
\end{eqnarray}
Once again we write down the matrix element of $y^k$ in the eigenstates of the Lewis invariant $I(t)$. This reads
\begin{eqnarray}
\langle n,m-n|y^k|n,m'-n\rangle&=&\sum_{r=0}^{k}\frac{\pi}{2^k}~{^{k}C_r}\delta_{m',m+2r-k}\nonumber\\
&\times& \lambda_n^2(\hbar^{1/2} \rho)^{2n+k+2}\sqrt{m!(m+2r-k)!}\nonumber\\
&\times& \int_0^\infty dw~w^{n-m-r+k}e^{-w} L_m^{(n-m)}(w) L_{m+2r-k}^{(n-m-2r+k)}(w).
\label{eqn85I}
\end{eqnarray}
Using Eqn.(\ref{eqn85}), we may evaluate the matrix element of $y$ and $y^2$ 
in the eigenstate of the Hamiltonian. We find, 
\begin{eqnarray}
_{H}\langle n,m-n|y|n,m'-n\rangle_{H}&=&
_{H}\langle n,m-n|y^k\mid_{\,k=1;\,r=0}|n,m'-n\rangle_{H}\nonumber\\&&+ _{H}\langle n,m-n|y^k\mid_{\,k=1;\,r=1}|n,m'-n\rangle_{H}
\nonumber\\
&=&-\frac{1}{2}
{(\rho\hbar^{1/2})}[\sqrt{m}e^{-i\Theta_{0,1}}\delta_{m',m-1}
+\sqrt{m+1}e^{i\Theta_{0,1}}\delta_{m',m+1}].
\label{eqn86}
\end{eqnarray}
\begin{eqnarray}
_{H}\langle n,m-n|y^2|n,m'-n\rangle_{H}&=&_{H}\langle n,m-n|y^k\mid_{\,k=2;\,r=0}|n,m'-n\rangle_{H}\nonumber\\
&&+_{H}\langle n,m-n|y^k\mid_{\,k=2;\,r=1}|n,m'-n\rangle_{H}\nonumber\\
&&+_{H}\langle n,m-n|y^k\mid_{\,k=2;\,r=2}|n,m'-n\rangle_{H} \nonumber\\
&=&\frac{\hbar\rho^2}{4}\delta_{m',m-2}\sqrt{m(m-1)}e^{-i\Theta_{0,2}}\nonumber\\
&&+\frac{1}{2}\delta_{m,m'}{(\hbar\rho^2)}(m+n+1)+\frac{\hbar\rho^2}{4}\delta_{m',m+2}\sqrt{(m+2)(m+1)}e^{i\Theta_{0,2}}.\nonumber\\
\label{eqn87}
\end{eqnarray}
From the above analysis, we find that even the expression for the matrix element of the operator $y^k$ in the eigenstate of $H(t)$ 
matches with that found in \cite{Dey} for $k=1, 2$, 
though again they had inappropriately quoted the results in the 
eigenstate of the Lewis invariant. 

%%%%%%%%%%%%%%%%%%%%%%%%%%%%%%%%%%%%%%%%%%%%%%%%%%%%%%%%%%%%%%%%%%%

%%%%%%%%%%%%%%%%%%%%%%%%%%%%%%%%%%%%%%%%%%%%%%%%%%%%%%%%%%%%%%%%%%%

\subsection{Analysis of the expectation value of energy}
As we have already seen from Eqn.(\ref{eqn2}), in order to calculate the expectation value of energy one needs the expectation values $\braket{{p_1}^2}$, $\braket{{p_2}^2}$, $\braket{{x_1}^2}$, $\braket{{x_2}^2}$, $\braket{{p_1}{x_2}}$ and 
$\braket{{p_2}{x_1}}$. As we have seen in the previous subsection, our calculated generalized expressions for matrix elements $_{H}\langle n,m-n|x^k\mid_{\,k=1;\,r=0}|n,m'-n\rangle_{H}$ and $_{H}\langle n,m-n|y^k\mid_{\,k=1;\,r=0}|n,m'-n\rangle_{H}$ matched exactly with the calculations in \cite{Dey} for $k=1, 2$. Hence, we use the matrix elements quoted in the said work to calculate 
the following expectation values,
\begin{eqnarray}
\braket{x_j^2}=\dfrac{\rho^2}{2}(n+m+1)\,\,;\,\,\braket{p_j^2}=\dfrac{1}{2}\left(\dfrac{1}{\rho^2}+\dfrac{\dot{\rho}^2}{a^2} \right)\,(n+m+1)\,\,;\,\,\braket{x_j\,p_k}=\dfrac{1}{2}\,\epsilon_{jk}(m-n)\,\,;\label{eqn88}
\end{eqnarray}
where $j,k=1,2$ and $\epsilon_{jk}=-\epsilon_{kj}$ with $\epsilon_{12}=1$. So, the expectation value of energy $\braket{E_{n,m-n}(t)}$ with 
respect to energy eigenstate $\psi_{n,m-n}(r,\theta,t)$ can be expressed as,
\begin{align}
&\braket{E_{n,m-n}(t)}=\dfrac{1}{2}\,(n+m+1)\left[b(t)\rho^2(t)+\dfrac{a(t)}{\rho^2(t)}+\dfrac{\dot{\rho}^2(t)}{a(t)} \right]+c(t)\,(n-m)\,\,.\nonumber\\
&=\dfrac{1}{2}\left[\,(n+m+1)\left(b(t)\rho^2(t)+\dfrac{a(t)}{\rho^2(t)}+\dfrac{\dot{\rho}^2(t)}{a(t)} \right)+(n-m)\left(\dfrac{f(t)\Omega(t)}{M}+\dfrac{M\omega^2(t)\theta(t)}{f(t)}\right) \right].
\label{eqn89}
\end{align}

%%%%%%%%%%%%%%%%%%%%%%%%%%%%%%%%%%%%%%%%%%%%%%%%%%%%%%%%%%%%%%%%%%%%%%%%%%%

%&=\dfrac{1}{2}\,(n+m+1)\left[b(t)\rho^2(t)+\dfrac{a(t)}{\rho^2(t)}+\dfrac{\dot{\rho}^2(t)}{a(t)} \right]+(n-m)\left[\omega(t)\sqrt{\dfrac{M\,a(t)}{f(t)}-1}+\sqrt{\dfrac{b(t)\,f(t)}{M}-\omega^2(t)} \right].

%%%%%%%%%%%%%%%%%%%%%%%%%%%%%%%%%%%%%%%%%%%%%%%%%%%%%%%%%%%%%%%%%%%%%%%%%%%

\noindent It is interesting to note that even when the frequency of oscillation $\omega{\rightarrow}0$, the 
expectation value of energy is non-zero. This is because all the three parameters of the Hamiltonian $a(t)$, $b(t)$ and $c(t)$ are finite even as $\omega{\rightarrow}0$, as is clear from the Eqn(s).(\ref{3},\ref{4},\ref{eqn3}). Now we will proceed to study the time-dependent behaviour of $\braket{E_{n,m-n}(t)}$ for various types of damping.

%%%%%%%%%%%%%%%%%%%%%%%%%%%%%%%%%%%%%%%%%%%%%%%%%%%%%%%%%%%%%%%%%

%%%%%%%%%%%%%%%%%%%%%%%%%%%%%%%%%%%%%%%%%%%%%%%%%%%%%%%%%%%%%%%%%

\subsubsection{Exponentially decaying solution}
%%%%%%%%%%%%%%%%%%%%%%%%%%%%%%%%%%%%%%%%%%%%%%%%%%%%%%%%%%%%%%%%%%%%%%%%%%%%%%%%%%%%%%%
For the exponentially decaying solution given by Eqn.(\ref{EPsoln1}), the energy expectation value takes the following form,
\begin{equation}
\braket{E_{n,m-n}(t)}=(n+m+1)\mu^2\Delta+c(t)\,(n-m)
\label{eqn90}
\end{equation}
where we have set the constant $\xi^2$ to unity and used the constraint relation given by Eqn.(\ref{EPreln1}).

\vskip 0.1cm

%%%%%%%%%%%%%%%%%%%%%%%%%%%%%%%%%%%%%%%%%%%%%%%%%%%%%%%%%%%%%

%%%%%%%%%%%%%%%%%%%%%%%%%%%%%%%%%%%%%%%%%%%%%%%%%%%%%%%%%%%%%

\noindent{\bf $\langle A\rangle$ Solution Set-Ia}
\vskip 0.15cm

\noindent For this case we consider $f(t)=1$ and $\omega(t)=\omega_0\,e^{-\Gamma\,t/2}$. The expectation value of 
energy for the ground state has the following expression,
\begin{eqnarray}
\braket{E_{n,-n}(t)}&=&(n+1)\mu^2\Delta+\,n\,\left[\sqrt{\dfrac{ \Delta\,exp\,(\Gamma{t})-M{\omega_0}^2\,exp\,(-\Gamma{t}) }{M}} 
\,\right.\nonumber\\
&&~~~~~~~~~~~~~~~~~~~~~~~~~~\left. +\,\omega_0\,exp\,(-\Gamma{t/2})\sqrt{\,M\sigma\,exp\,(-\Gamma{t})-1 \,}\right].
\label{eqn91}
\end{eqnarray}
From Eqn.(\ref{eqn91}), we see that the expectation value of the energy becomes complex beyond a certain time limit. The 
condition for getting the expectation value of energy to be real is as follows,
\begin{eqnarray}
M\,\sigma\,e^{-\Gamma\,t}>1\,\,\Rightarrow\,t\leq\,\dfrac{ln(M\,\sigma)}{\Gamma}~.\label{eqn92}
\end{eqnarray}

%%%%%%%%%%%%%%%%%%%%%%  Figure %%%%%%%%%%%%%%%%%%%%%%%%%%%

\begin{figure}[t]
\centering
\includegraphics[scale=0.4]{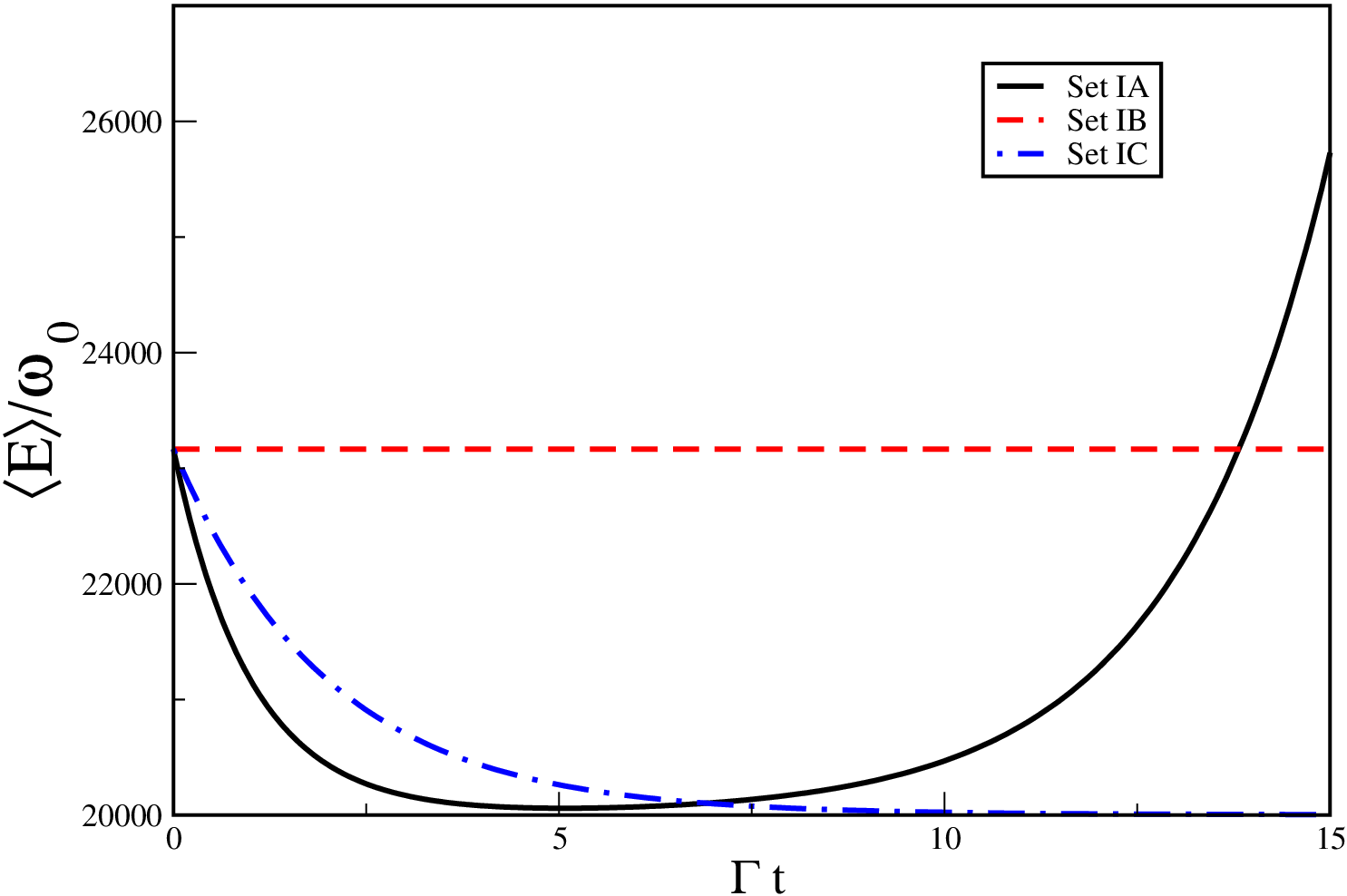}
\caption{\textit{A study of the variation of expectation value of energy, scaled by 
$\frac{1}{\omega_0}$ ($\frac{\langle E \rangle}{\omega_0}$) in order to make 
it dimensionless, as we vary $\Gamma$t (again a dimensionless quantity). Here 
we consider mass M=1, $\mu$=1,$\Delta$=$10^7$, $\sigma$=$10^7$, 
$\omega_0$=$10^3$ and $\Gamma$=1 in natural units. The expectation value of 
energy $\langle E \rangle$ is  calculated for exponentially decaying 
Hamiltonian parameters when $\langle A\rangle$ Set-IA $f(t)=1$ and 
$\omega(t)=\omega_0 e^{-{\Gamma}t/2}$; $\langle B\rangle$ Set-IB 
$f(t)=e^{-{\Gamma}t}$ and $\omega(t)=\omega_0$ and $\langle C\rangle$ 
Set-IC $f(t)=e^{-{\Gamma}t}$ and $\omega(t)=\omega_0 e^{-{\Gamma}t/2}$. While 
for $\langle A\rangle$ the energy first decreases, then increases with time, 
for $\langle B\rangle$ the energy remains constant as we vary time. 
For $\langle C\rangle$ the energy decays off with time.}}  
\label{fig1}
\end{figure}

%%%%%%%%%%%%%%%%%%%%%%%%%%%%%%%%%%%%%%%%%%%%%%%%%%%%%%%%%%

\noindent We see from Fig.(\ref{fig1}), that the energy initially decays but then increases with time. This is 
because for large time at which $exp\,(-\Gamma{t/2})\approx\,0 $, the 
approximated expression of energy reads
\begin{equation}
E_{n,-n}(t)\approx\,(n+1)\mu^2\Delta+\,n\,\sqrt{\dfrac{ \Delta\,exp\,(\Gamma{t}) }{M}}\,\,
\label{e7}
\end{equation}
which is still increasing with time. The reason for the increase of energy with time is the form of the 
coefficient $b(t)$ in the Hamiltonian. Although the coefficient $a(t)$ is exponentially decaying with 
time, the coefficient $b(t)$ exponentially increases with time in order to satisfy EP equation. However, since there is an upper limit of time within which the energy remains real, so the energy remains finite within the allowed time interval.

\vskip 0.1cm

%%%%%%%%%%%%%%%%%%%%%%%%%%%%%%%%%%%%%%%%%%%%%%%%%%%%%%%%%%%%%%%%%%%%%%%%%%%%%%%%%%%%%%%

%%%%%%%%%%%%%%%%%%%%%%%%%%%% Line No 1030 %%%%%%%%%%%%%%%%%%%%%%%%%%%%%%%%%%%%%%%%%%%%%

\noindent{{\bf{$\langle B\rangle$ Solution Set-Ib}}}

\noindent Here we set $f(t)=e^{-\Gamma\,t}$ and $\omega(t)=\omega_0$. With this the energy expression for the ground state takes the form,
\begin{equation}
\braket{E_{n,-n}(t)}=(n+1)\mu^2\Delta+\,n\,\left[\sqrt{\dfrac{ \Delta\,-M{\omega_0}^2\, }{M}} \,+\,\omega_0\sqrt{M\sigma-1}\right].\label{e8}
\end{equation}

%\textbf{Comparison with respect to damping case on commutative plane\,:-}\\
%Once again here is a notable point that the damped oscillator on NC plane can always possess a constant value of energy for this particular case of damping.This behaviour cannot be shown by a damped oscillator on commutative plane.

\noindent We note from Fig.(\ref{fig1}), that the expectation value of the energy remarkably remains constant as we vary time, as is observed from Eqn.(\ref{e8}). This must be because the effect of the exponentially decaying Hamiltonian 
coefficient $a(t)$ and damping term $f(t)$ gets balanced out by the exponentially increasing Hamiltonian 
coefficient $b(t)$.
\vskip 0.1cm

%%%%%%%%%%%%%%%%%%%%%%%%%%%%%%%%%%%%%%%%%%%%%%%%%%%%%%%%%%%%%%%%%%%

%%%%%%%%%%%%%%%%%%%%%%%%%%%%%%%%%%%%%%%%%%%%%%%%%%%%%%%%%%%%%%%%%%%

\noindent{{\bf{$\langle C\rangle$ Solution Set-Ic}}}

\noindent Here we set $f(t)=e^{-\Gamma\,t}$ and $\omega(t)=\omega_0\,e^{-\Gamma\,t/2}$. 
With this the expectation value of the energy expression takes the form,
\begin{equation}
\braket{E_{n,-n}(t)}=(n+1)\mu^2\Delta+\,n\,\left[\sqrt{\dfrac{ \Delta\,-M{\omega_0}^2\,exp\left[-\Gamma{t}\right] }{M}} 
 +  {\omega_0}\,exp\,(-\Gamma{t/2})\sqrt{M{\sigma}\,-1}\right].\label{e9}
\end{equation}
The above expression gives a very nice decaying expression for the expectation value of energy with respect to time, and finally approaching a constant value in the limit 
$t\rightarrow\infty$. This behaviour is also exhibited in the nature of the plot of variation of the expectation value of energy with time seen in Fig.(\ref{fig1}).

\vskip 0.1cm

%%%%%%%%%%%%%%%%%%%%%%%%%%%%%%%%%%%%%%%%%%%%%%%%%%%%%%%%%%%%%%%%

%%%%%%%%%%%%%%%%%%%%%%%%%%%%%%%%%%%%%%%%%%%%%%%%%%%%%%%%%%%%%%%%

\subsubsection{Rationally decaying solution}

\noindent In this case the expectation value of energy for $k=2$ reads
\begin{eqnarray}
E_{n,-n}(t)&=&\dfrac{(n+1)}{2(\Gamma\,t+\chi)}\left[2\left(\dfrac{\sigma}{\mu^2}+\Delta\mu^2\right)+\dfrac{\mu^2\Gamma^2}{8\sigma} \right]\nonumber\\
&&~~~~~~~~~~~~~~~~~~~~~~~ +n\left[\dfrac{\omega_0}{\Gamma\,t+\chi}\,\sqrt{\dfrac{4\sigma\,M}{(\Gamma{t}+\chi)^2}\,-\,1}\,+\,\sqrt{\dfrac{\Delta}{M}-\dfrac{\omega_0^{\,2}}{(\Gamma\,t+\chi)^{\,2}}} \right].\label{e10}
\end{eqnarray}
%\textbf{Comparison with respect to damping case on commutative plane\,:-}\\
\noindent Note that although it has a nice decaying property like the damping case on commutative plane, there is an upper bound of time above which the energy ceases to be real. The upper bound on time reads,
\begin{eqnarray}
4\sigma\,M\,\geq\,(\Gamma\,t+\chi)^2\,\,\Rightarrow\,\,t\,\leq\,\dfrac{1}{\Gamma}(2\sqrt{M\,\sigma}-\chi).
\label{e11}
\end{eqnarray}

%\pagebreak

\vskip 0.1cm
\begin{figure}[t]
\centering
\includegraphics[scale=0.4]{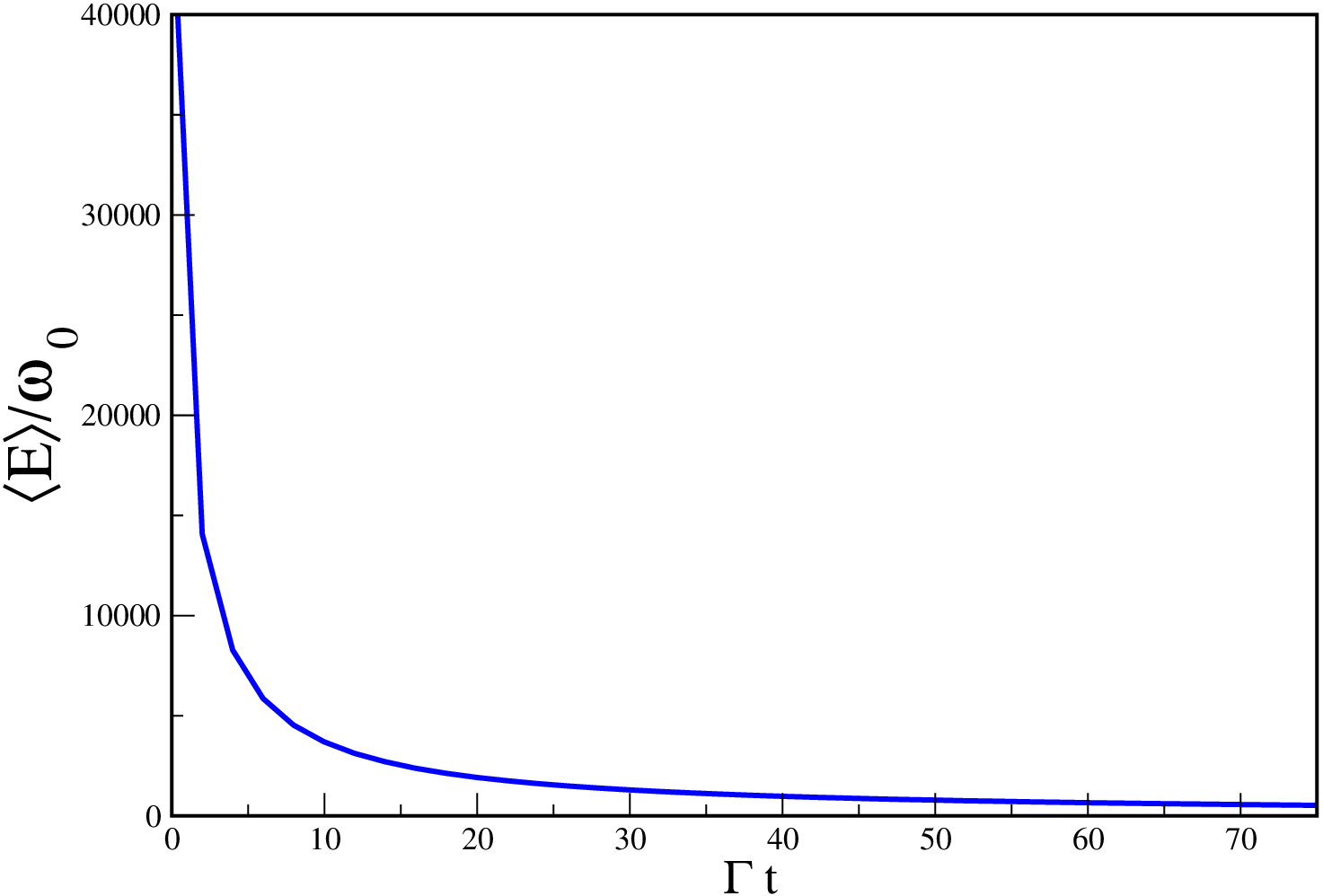}
\caption{\textit{A study of the variation of expectation value of energy, scaled by 
$\frac{1}{\omega_0}$ ($\frac{\langle E \rangle}{\omega_0}$) in order to make 
it dimensionless, as we vary $\Gamma$t (again a dimensionless quantity). Here 
we consider mass M=1, $\mu$=1,$\Delta$=$10^7$, $\sigma$=$10^7$, 
$\omega_0$=$10^3$, $\chi=1$ and $\Gamma$=1 in natural units. The expectation value of 
energy $\langle E \rangle$ is  calculated for rationally decaying 
Hamiltonian parameters. We consider $f(t)=1$ and $\omega(t)=\dfrac{\omega_0}{(\Gamma\,t+\chi)}$.}} 
\label{fig2}
\end{figure}

\noindent From Fig.(\ref{fig2}), we see indeed the expectation value of energy $\langle E \rangle$ decays with time 
following power law as expected for the rationally decaying solutions.  
%%%%%%%%%%%%%%%%%%%%%%%%%%%%%%%%%%%%%%%%%%%%%%%%%%%%%%%%%%%%%%%%%%%%

%%%%%%%%%%%%%%%%%%%%%%%%%%%%%%%%%%%%%%%%%%%%%%%%%%%%%%%%%%%%%%%%%%%%

\subsubsection{Elementary solution}
For the elementary solution set, the expectation value of the energy reads,
\begin{eqnarray}
\braket{E_{n,-n}(t)}&=&\dfrac{1}{2}(n+1)\left[\left(\Delta\mu^2+\dfrac{\sigma}{\mu^2}\right)\dfrac{1}{(\Gamma\,t+\chi)^2}+\dfrac{\mu^2\Gamma^2}{\sigma}\right]
\nonumber\\&&+n\left[\dfrac{\omega_0\sqrt{M\sigma-1}}{(\Gamma\,t+\chi)}+\dfrac{1}{(\Gamma\,t+\chi)}\sqrt{\dfrac{\Delta}{M\,(\Gamma\,t+\chi)^2}-\omega_0^2 } \right].\label{e12}
\end{eqnarray}
Further, the constraint relation $\Delta\mu^4=\xi^2\sigma$ results in the following form for the expectation value of energy (setting $\xi^2=1$),
\begin{align}
\braket{E_{n,-n}(t)}
&=\dfrac{1}{2}(n+1)\left[\dfrac{2\sigma}{\mu^2(\Gamma\,t+\chi)^2}+\dfrac{\mu^2\Gamma^2}{\sigma}\right]+n\,\left[\dfrac{\omega_0\sqrt{M\sigma-1}}{(\Gamma\,t+\chi)}+\dfrac{1}{(\Gamma\,t+\chi)}\sqrt{\dfrac{\Delta}{M\,(\Gamma\,t+\chi)^2}-\omega_0^2}\right].\label{e130}
\end{align}
%\textbf{Comparison with respect to damping case on commutative plane\,:-}\\
This expression also provides an upper bound of the time limit above which the expectation value of energy would become complex. This upper bound reads,
\begin{eqnarray}
\dfrac{\Delta}{M\,(\Gamma\,t+\chi)^2}\,\geq\,\omega_0^2\,\Rightarrow\,t\,\leq\,\dfrac{1}{\Gamma}\left[ \dfrac{1}{\omega_0}\sqrt{\dfrac{\Delta}{M}}-\chi \right].\label{e14}
\end{eqnarray}
%So, we have found some cases of dampimg where the the oscillation on NC plane behaves differently with respect to that on a commutative plane.
\begin{figure}[t]
\centering
\includegraphics[scale=0.4]{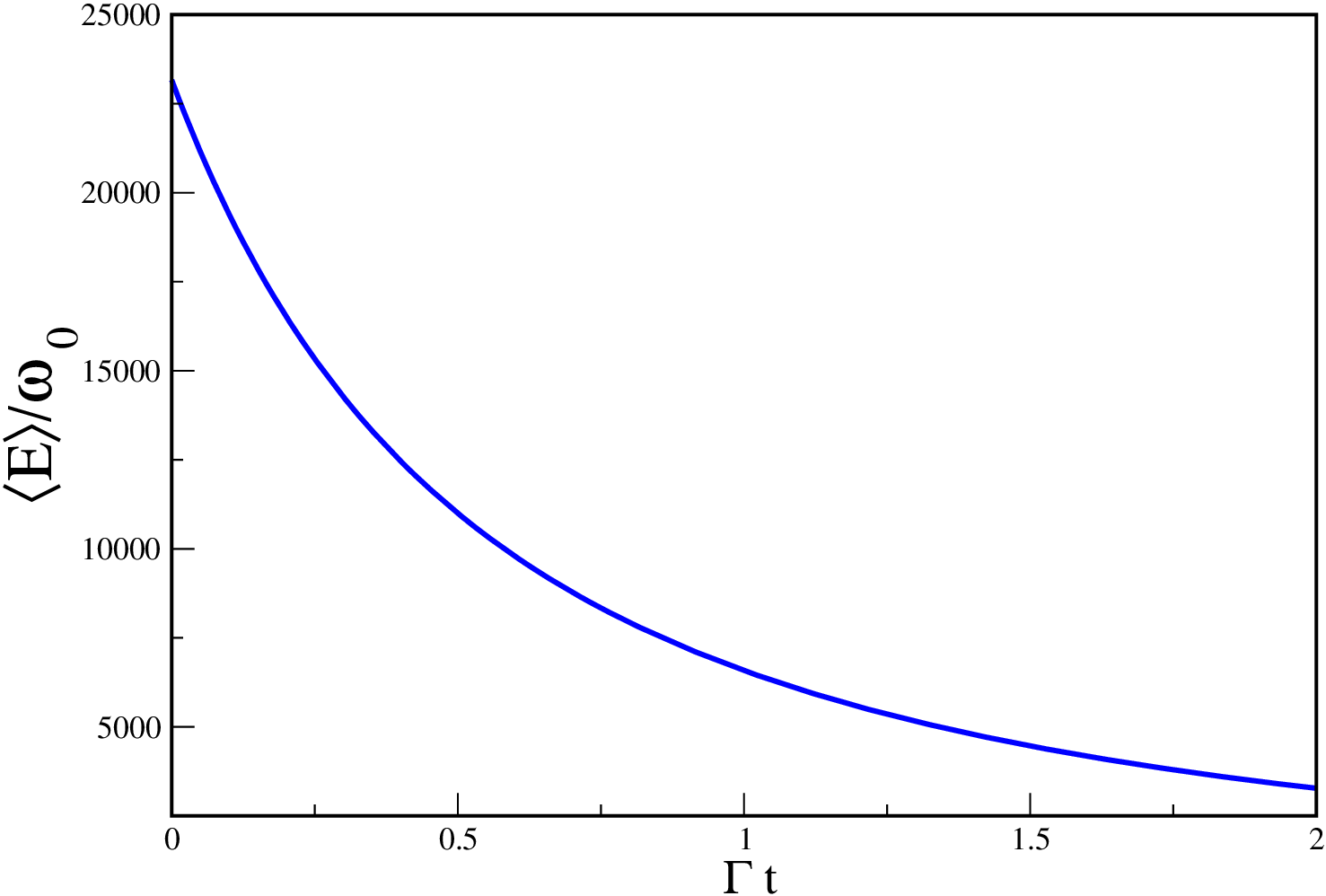}
\caption{\textit{A study of the variation of expectation value of energy, scaled by 
$\frac{1}{\omega_0}$ ($\frac{\langle E \rangle}{\omega_0}$) in order to make 
it dimensionless, as we vary $\Gamma$t (again a dimensionless quantity). Here 
we consider mass M=1, $\mu$=1,$\Delta$=$10^7$, $\sigma$=$10^7$, 
$\omega_0$=$10^3$, $\chi=1$ and $\Gamma$=1 in natural units. The expectation value of 
energy $\langle E \rangle$ is  calculated for elementarily decaying 
Hamiltonian parameters. We consider $f(t)=1$ and $\omega(t)=\dfrac{\omega_0}{(\Gamma\,t+\chi)}$.}} 
\label{fig3}
\end{figure}

\noindent In Fig.(\ref{fig3}), we observe that the expectation value of energy again undergoes a power law decay with time for the elementary solution.

%%%%%%%%%%%%%%%%%%%%%%%%%%%%%%%%%%%%%%%%%%%%%%%%%%%%%%%%%%%%%%%%%%%%%%%%%%%

%%%%%%%%%%%%%%%%%%%%%%%%%%%%%%%%%%%%%%%%%%%%%%%%%%%%%%%%%%%%%%%%%%%%%%%%%%%

%%%%%%%%%%%%%%%%%%%%%%%%%%%% Conclusion %%%%%%%%%%%%%%%%%%%%%%%%%%%%%%%%%%%%%%%%%%%%%%%

\section{Conclusion}
We now summarize our results. In this paper we have considered a two-dimensional damped harmonic oscillator in noncommutative space
with time dependent noncommutative parameters.
We map this system in terms of commutative variables by using a shift of variables connecting the noncommutative and commutative
space, known in the literature as Bopp-shift. We have then obtained the exact solution of this time dependent system by using the well known Lewis invariant which in turn leads to a non-linear differential equation known as the Ermakov-Pinney equation. We first obtain the Lewis invariant in Cartesian coordinates. 
We then make a transformation to polar coordinates and write down our results in these coordinates. Doing 
so, we use the operator approach to obtain the eigenstates of the invariant. With this background in place, we make various choices of the parameters in the problem which in turn leads to solutions for the time dependent noncommutative parameters. We have considered three different sets of choices for which solutions have been obtained, namely, exponentially decaying solutions, rationally decaying solutions and elementary solutions. Interestingly, the solutions obtained make it possible to integrate the phase factor exactly thereby giving an exact solution for the eigenstates of the Hamiltonian. We have then computed the matrix elements of operators raised to a finite integer power in both the eigenstates of the Hamiltonian as well as the Lewis invariant. From these results, we are able to compute the expectation value of the Hamiltonian. Expectedly, the expectation value of the energy varies with time. For the exponentially decaying solutions, we get three kinds of behaviour corresponding to the choices of the damping factor and the frequency of the oscillator. For the case where the damping factor is set to unity and the frequency of the oscillator decays with time, the expectation value of the energy first decreases with time and then increases.
The reason for this behaviour is due to the particular form of the solutions of the Ermakov-Pinney equation which fixes the forms of the noncommutative parameters. It is these time dependent forms of the noncommutative parameters that results in the above mentioned behaviour of the expectation value of the energy with time. 
In this case, we also observe that there is an upper bound of time above which the energy expectation value ceases to be real. 
For the case where the damping factor has a decaying part and the frequency of the oscillator is a constant, we observe that the expectation value of the energy remarkably remains constant with time. This must be the case because the effect of the exponentially decaying coefficient in the Hamiltonian 
and the damping term gets balanced out by the exponentially increasing coefficient in the Hamiltonian. For the case where both the damping term as well as the frequency of the oscillator decays with time, we find an exponentially decaying behaviour of the expectation value of the energy. For the rationally decaying and the elementary solution, we observe a power law decay of the energy expectation value with time together with an upper bound of time above which the energy expectation value ceases to be real. Investigating these cases of damped oscillators, we conclude that the behaviour corresponding to the exponentially decaying solution, where both the frequency and damping term are decaying exponentially with time, is similar to a damped oscillator in commutative space.

%%%%%%%%%%%%%%%%%%%%%%%%%%%%%%%%%%%%%%%%%%%%%%%%%%%%%%%%%%%%%%%%%%%%%%%%%%%%%%%%%%%%%%

%%%%%%%%%%%%%%%%%%%%%%%%%%%%%%%%%%% Appendix %%%%%%%%%%%%%%%%%%%%%%%%%%%%%%%%%%%%%%%%%

%\section{Appendix}

%%%%%%%%%%%%%%%%%%%%%%%%%%%%%%%%%%%%%%%%%%%%%%%%%%%%%%%%%%%%%%%%%%%%%%%%%%%%%%%%%%%%%%

%%%%%%%%%%%%%%%%%%%%%%%%%%%%%%%%%%%%%%%%%%%%%%%%%%%%%%%%%%%%%%%%%%%%%%%%%%%%%%%%%%%%%%

\section*{Acknowledgement}
MD would like to thank Ms. Riddhi Chatterjee and Ms.Rituparna Mandal for their helpful assistance to operate the software Mathematica.

%%%%%%%%%%%%%%%%%%%%%%%%%%%%%%%%%%%%%%%%%%%%%%%%%%%%%%%%%%%%%%%%%%%%%%%%%%%%%%%%%%%%%%%

%%%%%%%%%%%%%%%%%%%%%%%%%%%%%%%%%%%%%%%%%%%%%%%%%%%%%%%%%%%%%%%%%%%%%%%%%%%%%%%%%%%%%%%

%%%%%%%%%%%%%%%%%%%%%%%%%%%%%%%%%%%%%%%%%%%%%%%%%%%%%%%%%%%%%%%%%%%%%%%%%%%%%%%%%%%%%%%

%%%%%%%%%%%%%%%%%%%%%%%%%%%%%%%% References %%%%%%%%%%%%%%%%%%%%%%%%%%%%%%%%%%%%%%%%%%%

%%%%%%%%%%%%%%%%%%%%%%%%%%%%%%%%%%%%%%%%%%%%%%%%%%%%%%%%%%%%%%%%%%%%%%%%%%%%%%%%%%%%%%%%%%
%%%%%%%%%%%%%%%%%%%%%%%%%%%%%%%%%%%%%%%%%%%%%%%%%%%%%%%%%%%%%%%%%%%%%%%%%%%%%%%%%%%%%%%%%%%

\end{document}